\documentclass[twocolumn,showpacs,preprintnumbers,amsmath,amssymb,aps,pra,groupedaddress,showpacs]{revtex4}
\usepackage[colorlinks,citecolor=blue,linkcolor=red,hyperindex,CJKbookmarks]{hyperref}

\usepackage{amsmath}
\usepackage{graphicx}
\usepackage{dcolumn}

\usepackage{mathrsfs}
\usepackage{amssymb}
\usepackage{amsfonts}

\begin{document}
\title{Deterministic Entanglement Swapping in a Superconducting Circuit}

\author{Wen Ning$^{1}$}
\author{Xin-Jie Huang$^{1}$}
\author{Pei-Rong Han$^{1}$}
\author{Hekang Li$^{2}$}
\author{Hui Deng$^{4,5}$}
\author{Zhen-Biao Yang$^{1}$}
\email{zbyang@fzu.edu.cn}
\author{Zhi-Rong Zhong$^{1}$}
\author{Yan Xia$^{1}$}
\author{Kai Xu$^{2,3}$}
\email{kaixu@iphy.ac.cn}
\author{Dongning Zheng$^{2,3}$}
\author{Shi-Biao Zheng$^{1}$}
\email{t96034@fzu.edu.cn}
\affiliation{1.Fujian Key Laboratory of Quantum Information and Quantum Optics, College of Physics and Information Engineering, Fuzhou University, Fuzhou, Fujian 350108, China}
\affiliation{2.Institute of Physics and Beijing National Laboratory for Condensed Matter Physics, Chinese Academy of Sciences, Beijing 100190, China}
\affiliation{3.CAS Center for Excellence in Topological Quantum Computation, University of Chinese Academy of Sciences, Beijing 100190, China}
\affiliation{4.Hefei National Laboratory for Physical Sciences at Microscale and Department of Modern Physics, University of Science and Technology of China, Hefei, Anhui 230026, China}
\affiliation{5.CAS Centre for Excellence and Synergetic Innovation Centre in Quantum Information and Quantum Physics, University of Science and Technology of China, Hefei, Anhui 230026, China}
\date{\today }
\begin{abstract}
Entanglement swapping, the process to entangle two particles without coupling them in any way, is one of the most striking manifestations of the quantum-mechanical nonlocal characteristic. Besides fundamental interest, this process has applications in complex entanglement manipulation and quantum communication. Here we report a high-fidelity, unconditional entanglement swapping experiment in a superconducting circuit. The measured concurrence characterizing the qubit-qubit entanglement produced by swapping is above 0.75, confirming most of the entanglement of one qubit with its partner is deterministically transferred to another qubit that has never interacted with it. We further realize delayed-choice entanglement swapping, showing whether two qubits previously behaved as in an entangled state or as in a separable state is determined by a later choice of the type of measurement on their partners. This is the first demonstration of entanglement-separability duality in a deterministic way.
\end{abstract}

\maketitle
\begin{figure}[t]
\centering
\includegraphics[width=3.2in]{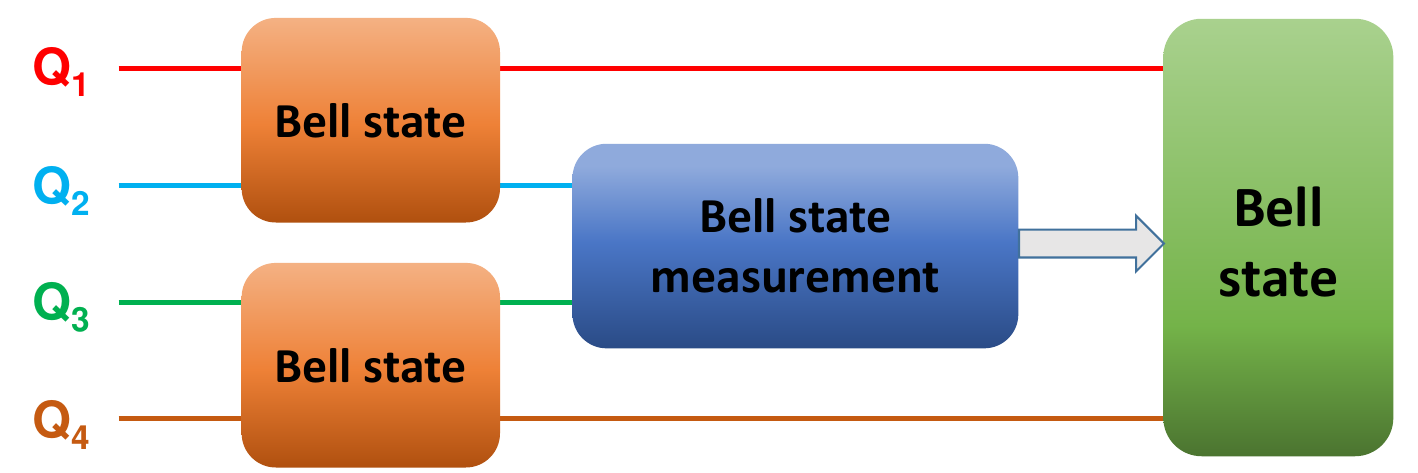}
\caption{Sketch of entanglement swapping. Initially, $Q_{1}$ is
entangled with $Q_{2}$ in a Bell state, and $Q_{3}$ with $Q_{4}$, but there
is no correlation between $Q_{1}$ and $Q_{4}$. A joint measurement on $Q_{2}$
and $Q_{3}$ in the Bell basis will project $Q_{1}$ and $Q_{4}$ to one of
four possible Bell states depending on the outcome of the $Q_{2}$-$Q_{3}
$ measurement.}
\label{f1}
\end{figure}

Quantum entanglement, lying at the heart of the Einstein-Podolsky-Rosen
(EPR) paradox \cite{einstein35}, is one of the most striking features of quantum
mechanics. When two particles are put in an entangled state, they can
exhibit nonlocal correlation that cannot be interpreted in terms of any
classical model as evidenced by violation of Bell's inequalities \cite{bell65,clauser69}. In
addition to fundamental tests of quantum mechanics, entanglement is an
essential resource for many quantum information tasks, such as quantum
teleportation \cite{bennett93} and measurement-based quantum computation \cite{raussendorf01}. The
nonlocal characteristic of quantum-mechanical wave functions allows two
particles that have never interacted to be put into an entangled state by
means of entanglement swapping \cite{zukowski93}. The process is illustrated in Fig. \ref{f1},
where the two qubits ($Q_{1}$ and $Q_{4}$) to be entangled are first
entangled with their respective partners ($Q_{2}$ and $Q_{3}$): $Q_{1}$ and $%
Q_{2}$ form the first entangled Bell pair, while $Q_{3}$ and $Q_{4}$ form
the second pair. Then a joint Bell state measurement applied to the partners
$Q_{2}$ and $Q_{3}$ will project the remaining two qubits, $Q_{1}$ and $%
Q_{4}$, to one of four possible Bell states; which entangled state is
produced depends on the outcome of the Bell state measurement. Aside from fundamental interest, entanglement swapping has
practical applications in quantum communication \cite{dur99} and in multipartite
entanglement manipulation necessary for construction of complex quantum
networks \cite{bose98}.

Entanglement swapping has been experimentally demonstrated with photonic
qubits \cite{pan98,pan01,riedmatten05,halder07,lu09,schmid09,sciarrino02,jennewein01,ma12}.
However, in these optical experiments, entanglement was
swapped conditional on the occurrence of preset photon coincidence events.
These events were detected only in a small fraction of experimental runs due
to the photon loss on optical components, lack of logic operations to
completely distinguish all four Bell states, and restriction of photon
detectors' efficiency \cite{ma12}. Experiments have realized heralded entanglement
between two spatially separated atomic qubits, each entangled with its
emitted photons before a partial Bell state analysis on these photons
\cite{matsukevich08,yuan08}; the entanglement was swapped also with a small probability. With
photonic continuous variables, unconditional entanglement swapping has been
reported~\cite{jia04,takei05}, but where only a small portion of entanglement was
preserved after swapping due to the limitation of the degree of entanglement
carried by the original entangled beams, which is the main source of the
infidelity for teleportation of a continuous-variable state \cite{braunstein98}.
Although unconditional teleportation has been demonstrated with different kinds of matter qubits, including nuclear magnetic resonance \cite{nielsen98},
trapped ions \cite{riebe04,barrett04}, and superconducting qubits \cite{baur12,steffen13}, where the teleported states were preset and
known to the experimenters, deterministic, high-fidelity entanglement swapping has only been realized in an ion trap \cite{riebe08}.
As pointed out in Ref. \cite{pan01},
entanglement swapping is the only known procedure that demonstrates the
quantum nature of teleportation$-$the qubit whose state is to be teleported
is entangled with another qubit, rendering it impossible to know this state.
When realized in a delayed-choice manner \cite{peres00}, this process reveals a more striking feature, that is,
one can \emph{a posteriori} determine two already detected particles previously behaved as an entangled pair or as a separable pair.

We here implement a deterministic entanglement swapping experiment with
superconducting qubits (labeled from $Q_{1}$ to $Q_{4}$).  Our results show that $Q_{1}$ and $Q_{4}$, although never coupled to each other,
are highly entangled after the swapping process, with the measured concurrence above 0.75. Unlike previous experiments with
photonic qubits, the Bell states are produced deterministically and the
measurement is single shot, so that the entanglement is swapped
unconditionally. We note that an entanglement swapping experiment also with superconducting qubits was briefly mentioned
in a recent review \cite{wendin}, but no experimental details have been released up to now. We further realize a delayed-choice entanglement swapping
experiment \cite{peres00}, where we choose to perform a Bell state measurement or a
separable-state measurement on $Q_{2}$ and $Q_{3}$ after $Q_{1}$ and $Q_{4}$
have been detected. The results demonstrate this later choice decides the
previous behavior of $Q_{1}$ and $Q_{4}$$-$whether they were entangled or
separable. This implies that entanglement is not a reality, but is a manifestation of the statistical correlation of the measured data;
the same set of data may show different types of correlations and have different interpretations when grouped in different manners.

\begin{figure}[t]
\centering
\includegraphics[width=3.2in]{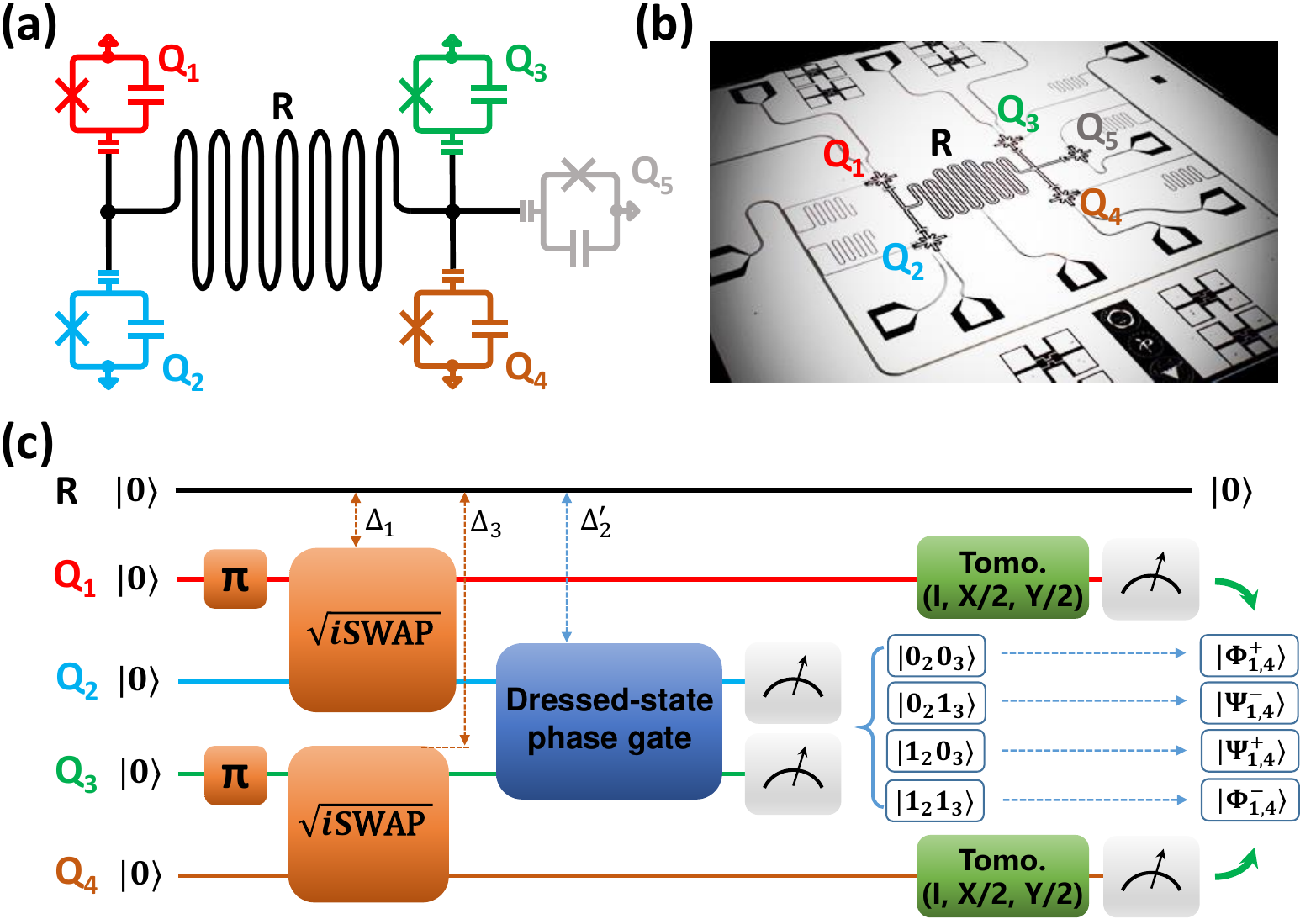}
\caption{Device schematic and pulse sequence. (a) Device
schematic. Five superconducting Xmon qubits (labeled from $Q_{1}$ to $%
Q_{5}$) are capacitively coupled to a bus resonator $R$. The frequency of each
qubit can be adjusted quickly, enabling the relevant qubit-resonator
interaction as well as the resonator-induced qubit-qubit couplings to be
effectively switched on and off. (b) Optical image of the device. (c)
Experimental sequence. The procedure consists of three parts: Generation of
Bell states for qubit pairs $Q_{1}$-$Q_{2}$ and $Q_{3}$-$Q_{4}$ via
resonator-mediated $\sqrt{\text{iSWAP}}$ gates following $\pi $ pulses
applied to $Q_{1}$ and $Q_{3}$; complete Bell state measurement on $Q_{2}$
and $Q_{3}$, achieved by subsequentially applying a dressed-state phase gate
on $Q_{2}$ and $Q_{3}$ and performing a joint detection in the basis $%
\left\{ \left\vert 0_{2}\right\rangle \left\vert 0_{3}\right\rangle
,\left\vert 0_{2}\right\rangle \left\vert 1_{3}\right\rangle ,\left\vert
1_{2}\right\rangle \left\vert 0_{3}\right\rangle ,\left\vert
1_{2}\right\rangle \left\vert 1_{3}\right\rangle \right\} $; 2-qubit quantum
state tomography for $Q_{1}$ and $Q_{4}$. The detailed pulse sequence is shown
in Fig. S3(a) of the Supplemental Material~\cite{suppl}.}
\label{f2}
\end{figure}

The device used to perform the experiment swapping is identical to that used
in Ref. \cite{song17_2}, where a resonator with a fixed frequency $\omega _{r}/2\pi= \times 5.588$ GHz is
controllably coupled to five superconducting Xmon qubits, whose
frequencies can be individually adjusted on nanosecond timescales using flux
bias lines. The device is sketched in Fig. \ref{f2}(a), and the optical image shown
in Fig. \ref{f2}(b). Throughout the experiment, $Q_{5}$ (unused) is tuned far
off resonance with the resonator and the other qubits, and will not be
included in the description of the system. The parameters of the system are
detailed in the Supplemental Material ~\cite{suppl}. All the qubits and the resonator are
initially in their ground states. The experiment starts with applying $\pi $
pulses to $Q_{1}$ and $Q_{3}$, transforming each of them from the ground
state $\left\vert 0\right\rangle $ to the excited state $\left\vert
1\right\rangle $ at its idle frequency, with the experimental sequence shown
in Fig. \ref{f2}(c). Then the qubit pairs $Q_{1}$-$Q_{2}$ and $Q_{3}$-$Q_{4}$ are
red detuned from the resonator by $\Delta _{1}=\Delta _{2}=2\pi \times 308$
MHz and $\Delta _{3}=\Delta _{4}=2\pi \times 238$ MHz, respectively. With this
setting, the resonator will not exchange photons with the qubits and remain in
the ground state, but it can simultaneously mediate two entangling gates,
each operating on one qubit pair \cite{zheng00,osnaghi01,majer07,zhong16,song17}, with the coupling between these two qubit
pairs being negligible owing to their large detuning~\cite{song17}.

The qubit pair, $Q_{j}$-$Q_{k}$ ($j=1$, $k=2$ or $j=3$, $k=4$),
evolves to the Bell state $\left\vert \Psi _{j,k}^{+}\right\rangle
=(\left\vert 1_{j}\right\rangle \left\vert 0_{k}\right\rangle +i\left\vert
0_{j}\right\rangle \left\vert 1_{k}\right\rangle )/\sqrt{2}$ after the
corresponding $\sqrt{\text{iSWAP}}$ gate.
As soon as $\left\vert \Psi _{j,k}^{+}\right\rangle $ is generated, $Q_{j}$
and $Q_{k}$ are detuned from each other to stop their coupling. The measured
density matrices for these two produced entangled pairs are displayed in the
Supplemental Material~\cite{suppl}. Their fidelities to the ideal Bell states are,
respectively, $F_{1,2}=0.982\pm0.006$ and $F_{3,4}=0.978\pm0.007$.
\begin{figure}[t]
\centering
\includegraphics[width=3.2in]{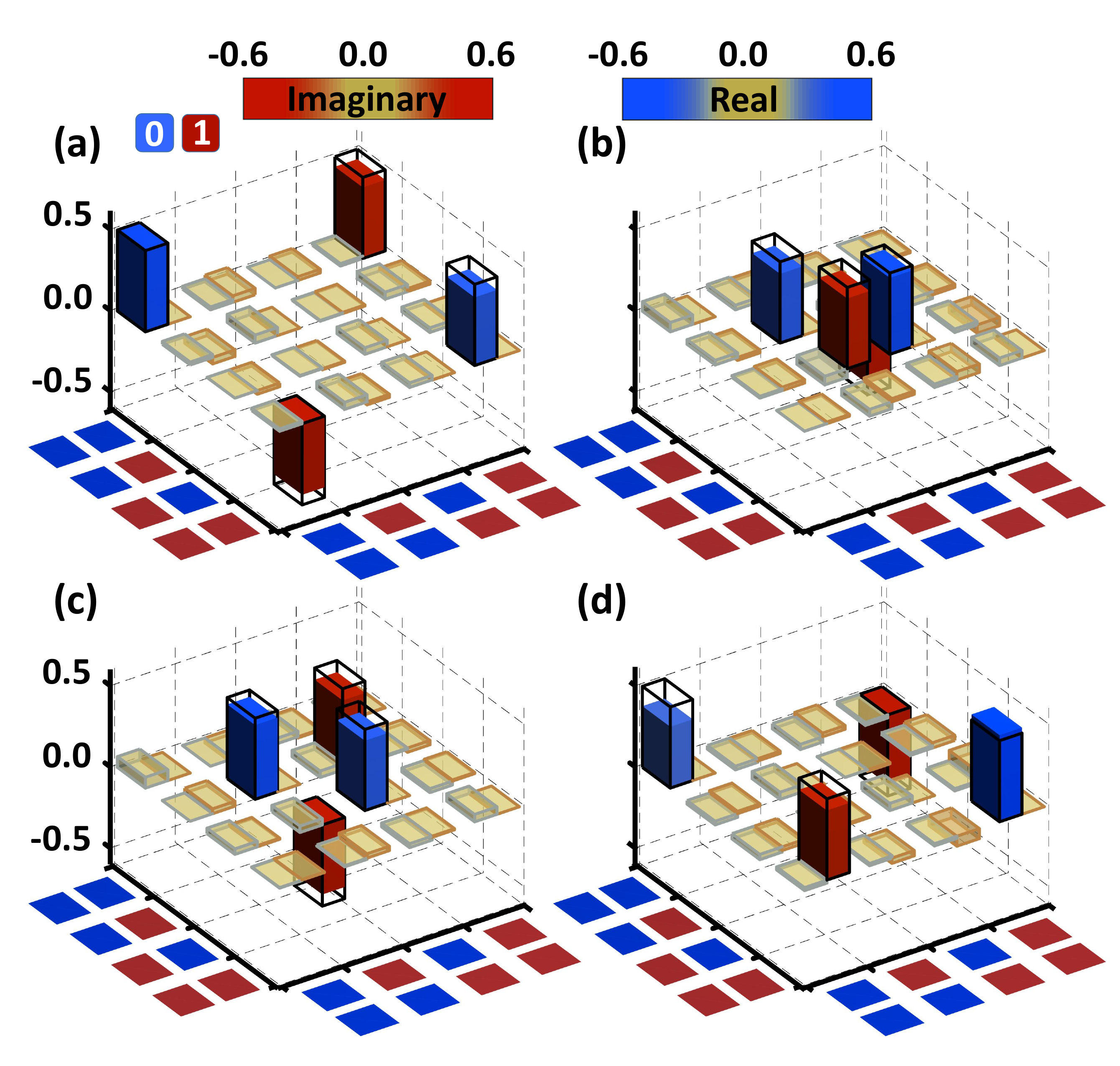}
\caption{Measured $Q_{1}$-$Q_{4}$ density matrices conditional
on the four $Q_{2}$-$Q_{3}$ measurement outcomes: (a) $\left\vert
0_{2}\right\rangle \left\vert 0_{3}\right\rangle $;  (b) $\left\vert
0_{2}\right\rangle \left\vert 1_{3}\right\rangle $; (c) $\left\vert
1_{2}\right\rangle \left\vert 0_{3}\right\rangle $; (d) $\left\vert
1_{2}\right\rangle \left\vert 1_{3}\right\rangle $. The results are obtained
with the experiment sequence shown in Fig. \ref{f2}(c). Each matrix element is
characterized by two color hbars, one for the real part and the other for the
imaginary part. The black wire frames denote the matrix
elements of the ideal output states.}
\label{f3}
\end{figure}
\begin{figure*}[t]
\centering
\includegraphics[width=6.4in]{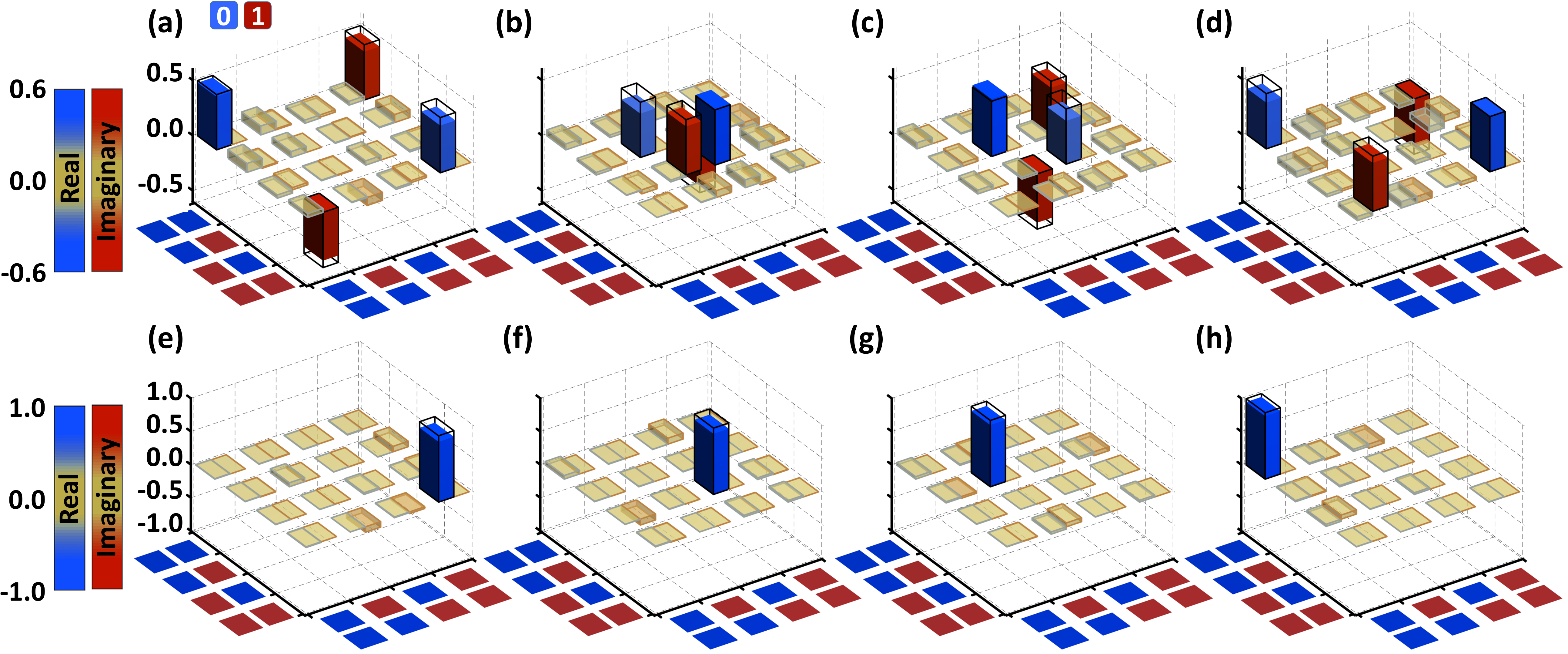}
\caption{Measured $Q_{1}$-$Q_{4}$ density matrices conditional
on outcomes of delayed-choice $Q_{2}$-$Q_{3}$ measurement. (a)-(d) Results obtained from the four subsets of data correlated with the
outcomes \{$\left\vert 0_{2}\right\rangle \left\vert 0_{3}\right\rangle $, $%
\left\vert 0_{2}\right\rangle \left\vert 1_{3}\right\rangle $, $\left\vert
1_{2}\right\rangle \left\vert 0_{3}\right\rangle $, $\left\vert
1_{2}\right\rangle \left\vert 1_{3}\right\rangle $\} of $Q_{2}$-$Q_{3}$
measurement performed after the dressed-state phase gate. Compared with Fig. \ref{f2}(c), the temporal
orders of $Q_{2}$-$Q_{3}$ Bell measurement and $Q_{1}$-$Q_{4}$
joint state tomography are inverted, with the experimental pulse sequence shown in
the Supplemental Material~\cite{suppl}. (e)-(h) Results obtained from the four
subsets of data correlated with the outcomes of the later $Q_{2}$-$Q_{3}$
measurement without the dressed-state phase gate.}
\label{f4}
\end{figure*}

The product of the two Bell states $\left| \Psi _{1,2}^{+}\right\rangle $
and $\left| \Psi _{3,4}^{+}\right\rangle $ can be expanded as
\begin{equation}
\begin{array}{c}
\left\vert \psi \right\rangle =\frac{1}{2}[-i\left\vert \Psi
_{2,3}^{+}\right\rangle \left\vert \Psi _{1,4}^{-}\right\rangle +i\left\vert
\Psi _{2,3}^{-}\right\rangle \left\vert \Psi _{1,4}^{+}\right\rangle  \\
+\left\vert \Phi _{2,3}^{+}\right\rangle \left\vert \Phi
_{1,4}^{+}\right\rangle -\left\vert \Phi _{2,3}^{-}\right\rangle \left\vert
\Phi _{1,4}^{-}\right\rangle ],%
\end{array}%
\end{equation}%
where $\left\vert \Psi _{j,k}^{\pm }\right\rangle =(\left\vert
1_{j}\right\rangle \left\vert 0_{k}\right\rangle \pm i\left\vert
0_{j}\right\rangle \left\vert 1_{k}\right\rangle )/\sqrt{2}$ and $\left\vert
\Phi _{j,k}^{\pm }\right\rangle =(\left\vert
1_{j}\right\rangle \left\vert 1_{k}\right\rangle \pm i\left\vert
0_{j}\right\rangle \left\vert 0_{k}\right\rangle )/\sqrt{2}$. To realize entanglement
swapping, we perform a measurement on $Q_{2}$ and $Q_{3}$ in
the Bell basis $\left\{ \left\vert \Psi _{2,3}^{+}\right\rangle ,\left\vert
\Psi _{2,3}^{-}\right\rangle ,\left\vert \Phi _{2,3}^{+}\right\rangle
,\left\vert \Phi _{2,3}^{-}\right\rangle \right\} $, which will project $%
Q_{1}$ and $Q_{4}$ to one Bell state. A complete Bell state measurement can
be implemented by mapping the Bell basis onto the computational basis $%
\left\{ \left\vert 0_{2}\right\rangle \left\vert 0_{3}\right\rangle
,\left\vert 0_{2}\right\rangle \left\vert 1_{3}\right\rangle ,\left\vert
1_{2}\right\rangle \left\vert 0_{3}\right\rangle ,\left\vert
1_{2}\right\rangle \left\vert 1_{3}\right\rangle \right\} $ through a
dressed-state phase gate \cite{guo18,song18}.
We note that the $\sqrt{\text{iSWAP}}$ gate only transforms two out of the four
Bell states into product states, and thus cannot be used for deterministically
distinguishing all the Bell states. To implement the dressed-state phase gate between $Q_{2}$ and $Q_{3}$,
we tune $Q_{1}$ and $Q_{4}$ back
to their idle frequencies, so that neither of them can interact with other
qubits, and then red detune $Q_{2}$ and $Q_{3}$ from the resonator by the
same amount $\Delta _{2}^{^{\prime }}=\Delta _{3}^{^{\prime }}=2\pi\times 308$ MHz,
switching on their interaction via the resonator-induced virtual photon
exchange, with the coupling strength $\lambda _{2,3}=2\pi\times1.14$ MHz. At the same time, we apply a resonant continuous drive
to each of these two qubits, whose phase is inverted in the middle of the
two-qubit interaction with a duration $\tau _{2,3}=\pi /2\lambda _{2,3}$.
When the difference of the Rabi frequencies of these two drives is much
larger than $\lambda _{2,3}$, a dressed-state phase gate between $Q_{2}$ and
$Q_{3}$ is achieved. As a result, the four Bell states of $Q_{2}$ and $Q_{3}$
evolve as (see Supplemental Material~\cite{suppl})
\begin{eqnarray}
\left\vert \Psi _{2,3}^{+}\right\rangle  &\rightarrow &i\left\vert
0_{2}\right\rangle \left\vert 1_{3}\right\rangle ,  
\left\vert \Psi _{2,3}^{-}\right\rangle  \rightarrow \left\vert
1_{2}\right\rangle \left\vert 0_{3}\right\rangle ,  \nonumber \\
\left\vert \Phi _{2,3}^{+}\right\rangle  &\rightarrow &i\left\vert
0_{2}\right\rangle \left\vert 0_{3}\right\rangle ,  
\left\vert \Phi _{2,3}^{-}\right\rangle  \rightarrow \left\vert
1_{2}\right\rangle \left\vert 1_{3}\right\rangle .
\end{eqnarray}%
The combination of this transformation and the subsequent detection of $Q_{2}
$ and $Q_{3}$ in the computational basis $\{\left\vert
0_{2}\right\rangle \left\vert 0_{3}\right\rangle ,\left\vert
0_{2}\right\rangle \left\vert 1_{3}\right\rangle ,\left\vert
1_{2}\right\rangle \left\vert 0_{3}\right\rangle ,\left\vert
1_{2}\right\rangle \left\vert 1_{3}\right\rangle \}$
effectively realizes the complete
Bell state analysis, enabling us to distinguish all the four Bell states.
Consequently, $Q_{1}$ and $Q_{4}$ are randomly projected onto one of the
four Bell states $\left\{ \left\vert \Phi _{1,4}^{+}\right\rangle
,\left\vert \Psi _{1,4}^{-}\right\rangle ,\left\vert \Psi
_{1,4}^{+}\right\rangle ,\left\vert \Phi _{1,4}^{-}\right\rangle \right\} $
depending on the $Q_{2}$-$Q_{3}$ measurement outcome.
During the $Q_{2}$-$Q_{3}$ Bell-state measurement, $Q_{1}$ and $Q_{4}$ respectively stay at their idle frequencies,
so that the interactions of each of them with the resonator and with any other qubit are effectively switched off due to the large detunings.

After the Bell state analysis, we perform joint 2-qubit state tomography to
reconstruct the density matrix for $Q_{1}$ and $Q_{4}$. The measured density
matrices of $Q_{1}$ and $Q_{4}$ conditional on the measurement outcomes $%
\left\vert 0_{2}\right\rangle \left\vert 0_{3}\right\rangle $, $\left\vert
0_{2}\right\rangle \left\vert 1_{3}\right\rangle $, $\left\vert
1_{2}\right\rangle \left\vert 0_{3}\right\rangle $, and $\left\vert
1_{2}\right\rangle \left\vert 1_{3}\right\rangle $ of $Q_{2}$ and $Q_{3}$
are displayed in Figs. \ref{f3}(a)-(d), respectively. The readout error of each qubit is corrected when reconstructing these density matrices.
Ideally, for these four outcomes $Q_{1}$ and $%
Q_{4}$ are projected onto $\left\vert \Phi _{1,4}^{+}\right\rangle $, $%
\left\vert \Psi _{1,4}^{-}\right\rangle, ${\bf \ }$\left\vert \Psi
_{1,4}^{+}\right\rangle $, and $\left\vert \Phi _{1,4}^{-}\right\rangle $,
respectively. The fidelities for the four obtained Bell states to the ideal
ones are $F_{\Phi ^{+}}=0.893\pm0.010$, $F_{\Psi ^{-}}=0.879\pm0.010$, $F_{\Psi ^{+}}=0.872\pm0.011$, and $%
F_{\Phi ^{-}}=0.884\pm0.010$, with the concurrences $C_{\Phi ^{+}}=0.794\pm0.020$, $C_{\Psi ^{-}}=0.779\pm0.020$%
, $C_{\Psi ^{+}}=0.758\pm0.024$, and $C_{\Phi ^{-}}=0.785\pm0.021$, respectively. These results show
that $Q_{4}$ ($Q_{1}$) inherits most of the entanglement of $Q_{2}$ with $%
Q_{1}$ ($Q_{3}$ with $Q_{4}$) after the swapping, which is in stark contrast
with experiments with photonic continuous variables \cite{jia04,takei05}, where only a
small portion of entanglement is inherited (e.g., about 29\% in Ref. \cite{jia04}).
As the readout error of each qubit is corrected when reconstructing the $Q_1$-$Q_4$ output density matrices, the infidelities mainly come from imperfect preparation of the $Q_1$-$Q_2$ and $Q_3$-$Q_4$ Bell states, imperfection of the $Q_2$-$Q_3$ dressed-state phase gate, and decoherence effects of $Q_1$ and $Q_4$ during this gate.

We note that the deterministic entanglement swapping requires reliable Bell state measurement on $Q_2$ and $Q_3$, whose performance depends on the quality of the dressed-state phase gate and the single-shot state readout fidelities of $Q_2$ and $Q_3$. The fidelity of the dressed-state phase gate is $F_{gt}\approx0.966$, while the average readout fidelities of $Q_2$ and $Q_3$ are $F_2=0.95$ and $F_3=0.94$, where $F_j=(F_{0,j}+F_{1,j})/2$, with $F_{0,j}$ and $F_{1,j}$ denoting the $\left\vert0\right\rangle$- and $\left\vert1\right\rangle$-state readout fidelities of $Q_j$, whose values are listed in Table S1 of the Supplemental Material~\cite{suppl}. Without readout error corrections, the average measurement fidelity of the four Bell states is roughly $F_{gt} F_2 F_3\approx0.863$, and each of the four corresponding $Q_1$-$Q_4$ output states, as shown in Fig. S4 of the Supplemental Material~\cite{suppl}, has a fidelity above $0.76$ and a concurrence exceeding $0.54$.

Going one step further, we delay the $Q_{2}$-$Q_{3}$ Bell state measurement
until the joint $Q_{1}$-$Q_{4}$ state has been detected. The detailed pulse
sequence is shown in Fig. S3(b) of the Supplemental Material~\cite{suppl},
where the $Q_{2}$-$Q_{3}$ readout pulse is applied about $219$ ns after the end of $Q_{1}$-$Q_{4}$ readout pulse.
Since the correlation between the outcomes of $Q_{2}$-$Q_{3}$
measurement and $Q_{1}$-$Q_{4}$ measurement is independent of their temporal
order, this arrangement will result in entanglement swapping in a delayed
manner \cite{peres00}.
According to $Q_{2}$-$Q_{3}$ Bell state measurement outcomes, the data of $%
Q_{1} $-$Q_{4}$ joint state measurement are sorted into four subsets, from
which four density matrices are reconstructed, and shown in Figs. \ref{f4}(a)-(d). As in the nondelayed case, these four density matrices correspond to four Bell states, with the respective
fidelities $F_{\Phi ^{+}}$=$0.891\pm0.012$, $F_{\Psi ^{-}}$=$0.891\pm0.012$, $F_{\Psi ^{+}}$=$0.896\pm0.010$, and $%
F_{\Phi ^{-}}$=$0.897\pm0.010$, and concurrences $C_{\Phi ^{+}}$=$0.815\pm0.026$, $C_{\Psi ^{-}}$=$0.816\pm0.024$, $%
C_{\Psi ^{+}}$=$0.806\pm0.022$, and $C_{\Phi ^{-}}$=$0.807\pm0.019$. The fidelities and concurrences are
slightly higher than those in the nondelayed case due to the fact that $%
Q_{1}$-$Q_{4}$ joint state is detected earlier so that the measured data are
less affected by decoherence effects.

We also perform another experiment, where we choose to measure $Q_{2}$ and $%
Q_{3}$ in the computational basis (without performing the dressed-state
phase gate before detection of their states). Again, this measurement is
performed after $Q_{1}$-$Q_{4}$ joint state detection, with the pulse
sequence shown in Fig. S3(c) of the Supplemental Material~\cite{suppl}. The density matrices
reconstructed from the four subsets of $Q_{1}$-$Q_{4}$ measurement data,
each associated with one of $Q_{2}$-$Q_{3}$ measurement outcomes $%
\{\left\vert 0_{2}\right\rangle \left\vert 0_{3}\right\rangle ,\left\vert
0_{2}\right\rangle \left\vert 1_{3}\right\rangle ,\left\vert
1_{2}\right\rangle \left\vert 0_{3}\right\rangle ,\left\vert
1_{2}\right\rangle \left\vert 1_{3}\right\rangle \}$, are presented in Figs. \ref{f4}(e)-(h), respectively. As expected, these matrices correspond to
product states $\{\left\vert 1_{1}\right\rangle \left\vert
1_{4}\right\rangle ,\left\vert 1_{1}\right\rangle \left\vert
0_{4}\right\rangle ,\left\vert 0_{1}\right\rangle \left\vert
1_{4}\right\rangle ,\left\vert 0_{1}\right\rangle \left\vert
0_{4}\right\rangle \}$ with the fidelities $%
\{0.907\pm0.011, 0.914\pm0.009, 0.930\pm0.009, 0.949\pm0.008\}$. The concurrence
associated with each of these reconstructed matrices is approximate to 0 (see Table S4 of the Supplemental Material~\cite{suppl}).

The above results demonstrate whether or not the already measured qubits $%
Q_{1}$ and $Q_{4}$ previously behaved as an entangled pair depends on the
later choice of the type of measurement on $Q_{2}$ and $Q_{3}$. As a
generalization of Wheeler's delayed-choice experiment proposed for
illustrating the wave-particle duality of a single particle \cite{wheeler}, the
delayed-choice entanglement swapping experiment reveals the
entanglement-separability duality of two particles \cite{brukner05}. A realization of
this gedanken experiment was previously reported with photonic qubits \cite{ma12},
but where only two out of four basis states could be distinguished in each
of the two mutually exclusive measurements, so that the
entanglement-separability duality was only partially demonstrated: Whether
the $Q_{1}$-$Q_{4}$ states associated with the two indistinguishable $Q_{2}$-%
$Q_{3}$ basis states manifested a quantum or a classical correlation could
not be confirmed. Another problem is only a small fraction of events
coinciding with the distinguishable basis states was detected owing to the
photon loss on optical components (only 4.4\% photons left) and nonunity
photon detection efficiency.

We have demonstrated deterministic entanglement swapping with superconducting qubits
controllably coupled to a resonator. The qubit-qubit couplings mediated by the resonator
allows for both the controlled generation of the Bell states and complete Bell state analysis.
We have further deterministically realized delayed-choice entanglement swapping, demonstrating
whether two qubits exhibited entangled or separable behavior can be \emph{a posteriori} decided after
they have been measured. Our results indicate quantum entanglement of two quantum systems is a
manifestation of the statistical correlations of the measured data, instead of a reality.

We thank Haohua Wang at Zhejiang University for technical support. This work was supported by the National Natural Science Foundation
of China (Grants No. 11674060, No. 11874114, and No. 11875108), and the Strategic
Priority Research Program of Chinese Academy of Sciences (Grant No. XDB28000000).

\end{document}


\title{Supplemental Material for {\textquotedblleft}Deterministic entanglement swapping in a superconducting circuit{\textquotedblright}}

\author{Wen Ning$^{1}$}
\author{Xin-Jie Huang$^{1}$}
\author{Pei-Rong Han$^{1}$}
\author{Hekang Li$^{2}$}
\author{Hui Deng$^{4,5}$}
\author{Zhen-Biao Yang$^{1}$}
\email{zbyang@fzu.edu.cn}
\author{Zhi-Rong Zhong$^{1}$}
\author{Yan Xia$^{1}$}
\author{Kai Xu$^{2,3}$}
\email{kaixu@iphy.ac.cn}
\author{Dongning Zheng$^{2,3}$}
\author{Shi-Biao Zheng$^{1}$}
\email{t96034@fzu.edu.cn}
\affiliation{1.Fujian Key Laboratory of Quantum Information and Quantum Optics, College of Physics and Information Engineering, Fuzhou University, Fuzhou, Fujian 350108, China}
\affiliation{2.Institute of Physics and Beijing National Laboratory for Condensed Matter Physics, Chinese Academy of Sciences, Beijing 100190,China}
\affiliation{3.CAS Center for Excellence in Topological Quantum Computation, University of Chinese Academy of Sciences, Beijing 100190, China}
\affiliation{4.Hefei National Laboratory for Physical Sciences at Microscale and Department of Modern Physics, University of Science and Technology of China, Hefei, Anhui 230026, China}
\affiliation{5.CAS Centre for Excellence and Synergetic Innovation Centre in Quantum Information and Quantum Physics, University of Science and Technology of China, Hefei, Anhui 230026, China}

\date{\today}
\maketitle
\tableofcontents

\section{Device parameters}

The device used to perform the present experiment is the same as that
reported in Ref. \cite{song2017}, where five frequency-tunable superconducting Xmon
qubits, labeled from $Q_{1}$ to $Q_{5}$, are capacitively coupled to a common resonator. In our experiment, $Q_{5}$ is not used.
The full Hamiltonian of the system can be described as

\begin{eqnarray}
H&=& \hbar\omega_r a^{+}a + \hbar\sum_{j=1}^{5}\omega_{j}S_{j}^+S_{j}^-
+\hbar\sum_{j=1}^{5}g_{j}(S _{j}^{+}a+S _{j}^{-}a^{+}) \nonumber\\
&+&\hbar\sum_{j,k}^{}\lambda^{c}_{j,k}(S^{+}_{j}S^{-}_{k}+S^{+}_{k}S^{-}_{j}).
\label{Seq:no1}
\end{eqnarray}

Qubit frequencies $\omega_j/2\pi$ are individually tunable from 5 to 6 GHz while the resonator frequency $\omega_r/2\pi$ is fixed at about 5.588 GHz.
$g_{j}$ is the coupling strength between qubit $Q_j$ and the resonator with the magnitude listed in Table~\ref{table1}. By equally detuning the frequency of any two qubits far away from that of the resonator, we can realize the effective qubit-qubit interaction with the coupling strength of $g_j g_k/\Delta$ ($\Delta=\omega_j-\omega_r=\omega_k-\omega_r, \vert\Delta\vert\gg g_j,g_k$), which enables realizations of the $\sqrt{i\rm{SWAP}}$ gate and dressed-state phase gate used in the experiment. Note that except for the dominant resonator-mediated interaction, there exists very small direct couplings $\lambda^{c}_{j,k}$ in the system which have been reported elsewhere in similar devices \cite{10qGHZ,guo2018}.

Characterization of qubit performance is presented in Table~\ref{table1}. For technical details about the superconducting qubits, see Supplemental Material of Ref. \cite{10qGHZ}, which shares similar control methods to our experiment.

\begin{table*}[!htb]
	\centering
	\begin{tabular}{cccccccccc}
		\hline
		\hline
		&$\omega_{j}/2\pi$ (GHz)&$T_{1,j}$ ($\mu$s)&$T_{2,j}^*$ ($\mu$s)&$T_{2,j}^{\textrm{SE}}$ ($\mu$s)& $T_{\phi,j}^{DD}$ ($\mu$s)&$g_j/2\pi$ (MHz) &$1/\kappa^r_j (ns)$&$F_{0,j}$&$F_{1,j}$\\
		\hline
		$Q_1$&5.229&27.1&2.0&6.2&59.2&20.8&219&0.975&0.927\\
		$Q_2$&5.311&27.1&2.6&8.7&33.2&19.9&203&0.975&0.925\\
		$Q_3$&5.366&24.0&2.0&6.4&55.5&20.0&315&0.961&0.919\\
		$Q_4$&5.421&18.1&2.0&7.6&45.2&19.4&313&0.979&0.822\\
		\hline
		\hline
	\end{tabular}
	\caption{\label{table1} \textbf{Qubits characteristics.} $\omega_{j}/2\pi$ is the idle frequency of $Q_j$ where single-qubit rotation pulses and tomographic pulses are applied. $T_{1,j}$ and $T_{2,j}^*$ are the energy relaxation time and Ramsey dephasing time of $Q_j$ (Gaussian decay) respectively measured at the idle point. $T_{2,j}^{\textrm{SE}}$ is the dephasing time (Gaussian decay) with spin echo, while $T_{\phi,j}^{DD}$ denotes the dephasing time (exponential decay) under continuous driving for dynamical decoupling \cite{guo2018}. The continuous driving is used in our experiment as it protects the qubits from dephasing much more effectively compared with the spin-echo technique. $g_j$ is the coupling strength between $Q_j$ and the bus resonator, $\kappa_{j}^{r}$ represents the linewidth of the readout resonator for measuring the state of $Q_j$. $F_{0,j}$ ($F_{1,j}$) is the probability of detecting $Q_j$ in $\vert 0\rangle$ ($\vert 1\rangle$) when it is prepared in $\vert 0\rangle$ ($\vert 1\rangle$) state.}
\end{table*}

\section{Characterization of Bell states produced by $\protect\sqrt{i\rm{SWAP}}$ gates} 

\begin{figure}[!htb]
	\centering
	
	\includegraphics[width=0.49\textwidth,clip=True]{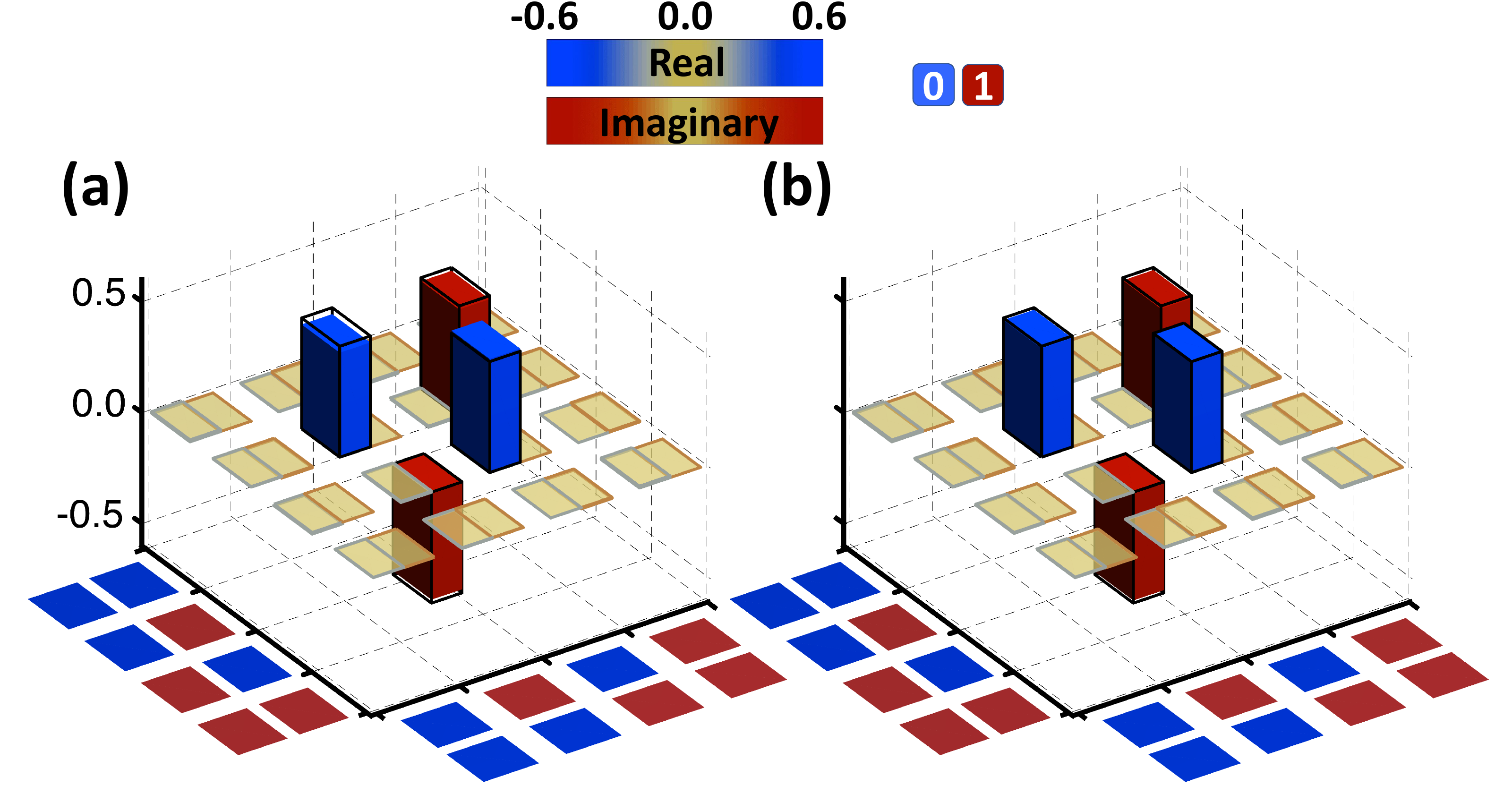}
	\caption{\label{seqbell} \footnotesize{Measured density matrices of Bell states produced
			via $\sqrt{i\rm{SWAP}}$ gates. (a) $Q_{1}$-$Q_{2}$ density matrix. (b) $%
			Q_{3}$-$Q_{4}$ density matrix. Each qubit pair is prepared in the state $%
			\left\vert 1\right\rangle \left\vert 0\right\rangle $ before the
			corresponding $\sqrt{i\rm{SWAP}}$ gate. The real and imaginary parts of each matrix element are separately characterized with two different color bars. The black wire frames represent the matrix elements of the ideal Bell states.
		}
	}
\end{figure}

The original $Q_{1}$-$Q_{2}$ and $Q_{3}$-$Q_{4}$ Bell states are produced
simultaneously \cite{10qGHZ}. The parallel operations for generating these entangled
states are realized by tuning the frequencies of these two qubit pairs to $5.28$
GHz and $5.35$ GHz, respectively. This frequency setting enables the resonator to
mediate two independent qubit-qubit swapping interactions, one between $Q_{1}
$ and $Q_{2}$ and the other between $Q_{3}$ and $Q_{4}$. We note that the
magnitude of the measured coupling $\lambda _{j,k}$ between qubits $Q_{j}$
and $Q_{k}$ is slightly smaller than the calculated resonator-induced
coupling. This is due to the fact that direct coupling
between these qubits with an opposite sign partly cancels out the
resonator-induced coupling \cite{10qGHZ,guo2018}. We characterize the Bell states generated
via the corresponding $\sqrt{i\rm{SWAP}}$ gates through joint state
tomography, with the $Q_{1}$-$Q_{2}$ and $Q_{3}$-$Q_{4}$ density matrices
displayed in Fig. \ref{seqbell}(a) and (b), respectively. The entanglements in the Bell
states are characterized by the magnitudes of the off-diagonal matrix elements
$\rho_{01,10}$ and $\rho_{10,01}$, which are about $0.49$ for both produced Bell pairs.
The fidelities of these produced Bell states are respectively $%
F_{1,2}=$0.982$\pm$0.006 and $F_{3,4}=$0.978$\pm$0.007. The populations of $\left\vert
0_{1}\right\rangle \left\vert 1_{2}\right\rangle \left\vert
0_{3}\right\rangle \left\vert 1_{4}\right\rangle $, $\left\vert
0_{1}\right\rangle \left\vert 1_{2}\right\rangle \left\vert
1_{3}\right\rangle \left\vert 0_{4}\right\rangle $, $\left\vert
1_{1}\right\rangle \left\vert 0_{2}\right\rangle \left\vert
0_{3}\right\rangle \left\vert 1_{4}\right\rangle $, and $\left\vert
1_{1}\right\rangle \left\vert 0_{2}\right\rangle \left\vert
1_{3}\right\rangle \left\vert 0_{4}\right\rangle $ in the joint computational basis of the
four qubits are 0.248$\pm$0.003, 0.241$\pm$0.006, 0.247$\pm$0.003, and 0.240$\pm$0.004, respectively. Slight deviations of these
populations from 1/4 are due to decoherence effects of these qubits and
limited detunings between each pair of qubits and the resonator and between
these two qubit pairs.

\begin{figure}[!htb]
	\centering
	\includegraphics[width=0.49\textwidth,clip=True]{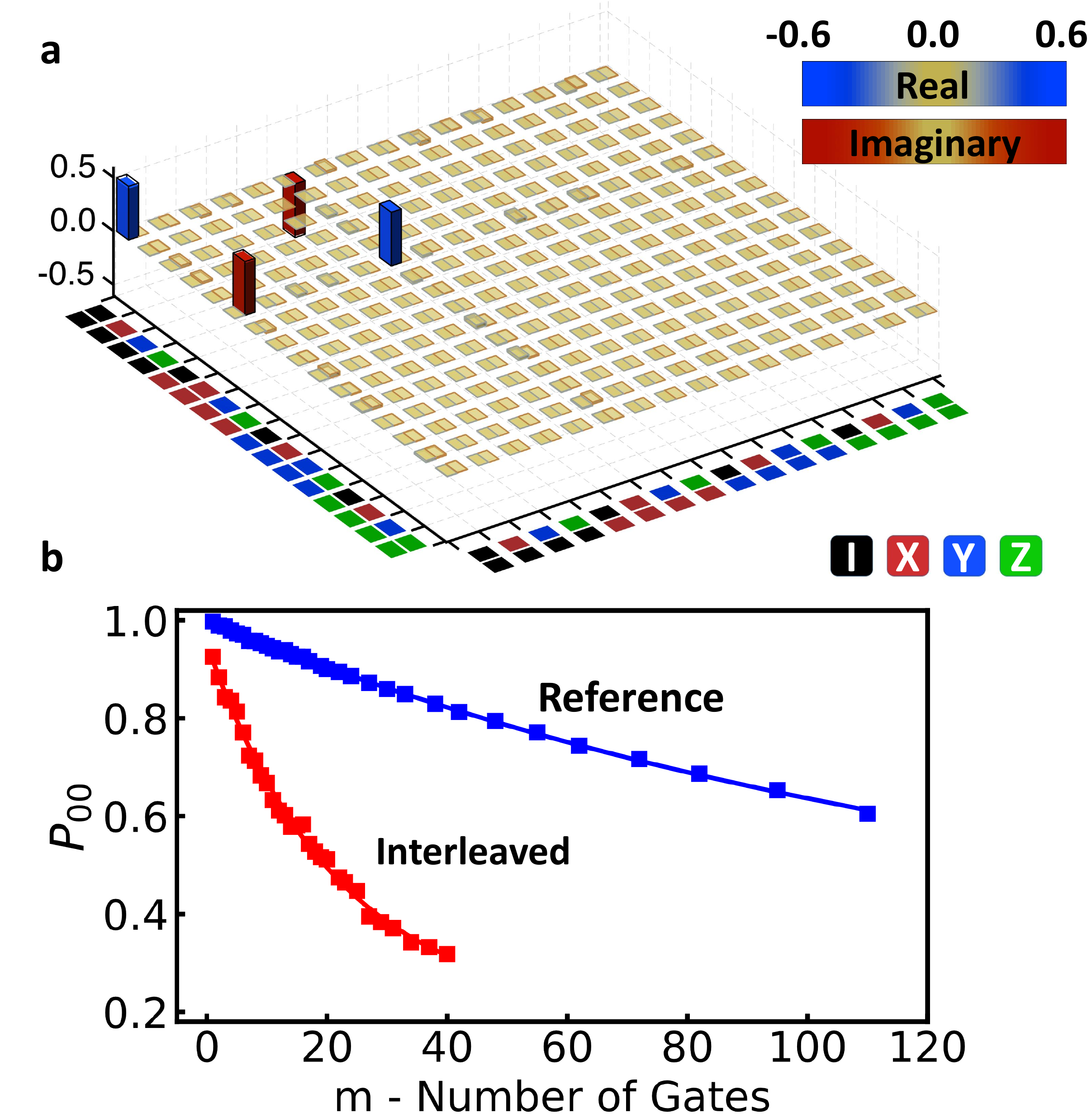}
	\caption{\label{seqchi} \footnotesize{Characterization of dressed-state phase gate. (a) Measured $\chi$ matrix  for the dressed-state phase gate in the Pauli basis. The real and imaginary parts of each element are plotted separately with different colorbars. The black wire frames represent the elements of ideal $\chi$. (b) Randomized benchmarking by inserting the dressed-state phase gate between random single-qubit Pauli gates. Plotted are the corrected probability of  $\vert 00\rangle$ state after a series of random Pauli gates with (red) and without (blue) the dressed-state phase gate inserted.
		}
	}
\end{figure}

\section{Complete Bell state measurement}

In our experiment, the $Q_2$-$Q_3$ Bell state measurement is enabled by a
dressed-state phase gate realized by a combination of  the resonator-mediated
$Q_2$-$Q_3$ swapping coupling and continuous resonant driving.
When $Q_{2}$ and $Q_{3}$ are red detuned from the resonator by the same
amount $\Delta _{2}^{^{\prime }}=\Delta _{3}^{^{\prime }}$ that is much
larger than the corresponding qubit-resonator couplings, the
resonator-induced $Q_{2}$-$Q_{3}$ coupling strength is $\lambda
_{2,3}=g_{2}g_{3}/\Delta _{2}^{^{\prime }}$ \cite{10qGHZ}. Under continuous driving,
the dynamics of $Q_{2}$ and $Q_{3}$ is described by the effective
Hamiltonian \cite{guo2018}
\begin{equation}
H_{\mathrm{eff}}=\hbar \left( -\lambda
_{2,3}S_{2}^{+}S_{3}^{-}+\sum_{j=2,3}\Omega _{j}e^{i\varphi
	_{j}}S_{j}^{+}\right) +H.c.,
\end{equation}%
where $\Omega _{j}$ and $\varphi _{j}$ denote the Rabi frequency and phase
of the drive applied to $Q_j$. We here assume that $\varphi
_{2}=\varphi _{3}=\varphi $. Under the condition $\left\vert \Omega
_{2}-\Omega _{3}\right\vert \gg \left\vert \lambda _{2,3}\right\vert $, the
effective Hamiltonian approximates
\begin{equation}
H_{\mathrm{eff}}^{^{\prime }}=-\frac{1}{2}\hbar \lambda_{2,3} S_{z,\varphi
	,2}S_{z,\varphi ,3}+\hbar \sum_{j=2,3}\Omega _{j}S_{z,\varphi ,j},
\end{equation}%
where $S_{z,\varphi ,j}=\left\vert +_{\varphi ,j}\right\rangle \left\langle
+_{\varphi ,j}\right\vert -\left\vert -_{\varphi ,j}\right\rangle
\left\langle -_{\varphi ,j}\right\vert $, with $\left\vert +_{\varphi
	,j}\right\rangle $ and $\left\vert -_{\varphi ,j}\right\rangle $ being the
dressed states, defined as $\left\vert +_{\varphi ,j}\right\rangle =(\left\vert 0_{j}\right\rangle +e^{i\varphi }\left\vert
1_{j}\right\rangle )/\sqrt{2} $ and $\left\vert -_{\varphi ,j}\right\rangle =(\left\vert 0_{j}\right\rangle -e^{i\varphi }\left\vert
1_{j}\right\rangle )/\sqrt{2}$. When the phase of each drive is inverted in the
middle of the pulse with the duration $\tau =\pi /2\lambda _{2,3}$, this
effective Hamiltonian leads to the controlled $\pi $-phase gate in the
dressed-state basis $\left\{ \left\vert +_{\varphi ,2}\right\rangle \left\vert
+_{\varphi ,3}\right\rangle ,\left\vert +_{\varphi ,2}\right\rangle
\left\vert -_{\varphi ,3}\right\rangle ,\left\vert -_{\varphi
	,2}\right\rangle \left\vert +_{\varphi ,3}\right\rangle ,\left\vert
-_{\varphi ,2}\right\rangle \left\vert -_{\varphi ,3}\right\rangle \right\} $
up to single-qubit operations $\exp (i\pi S_{z,\varphi ,j}/4)$. For
simplicity, we take $\varphi =0$. Then the evolution operator in the
computational basis $\left\{ \left\vert 0_{2}\right\rangle \left\vert
0_{3}\right\rangle ,\left\vert 0_{2}\right\rangle \left\vert
1_{3}\right\rangle ,\left\vert 1_{2}\right\rangle \left\vert
0_{3}\right\rangle ,\left\vert 1_{2}\right\rangle \left\vert
1_{3}\right\rangle \right\} $ is%
\begin{equation}
U=\frac{1}{\sqrt{2}}\left(
\begin{array}{cccc}
1 & 0 & 0 & i \\
0 & 1 & i & 0 \\
0 & i & 1 & 0 \\
i & 0 & 0 & 1%
\end{array}%
\right).
\end{equation}%
After the application of this evolution operator, each of the four Bell
states is transformed to a computational state, with the correspondence
given by Eq. (3) of the main text.

The dressed-state phase gate is characterized by both quantum process tomography and randomized benchmarking in experiment. The $\chi$ matrix obtained by quantum process tomography is shown in Fig. \ref{seqchi}, with a fidelity of 0.966$\pm$0.005, in agreement with that characterized by randomized benchmarking, which yields a fidelity of 0.971$\pm$0.002.

\begin{figure}[!htb]
	\centering
	\includegraphics[width=0.49\textwidth,clip=True]{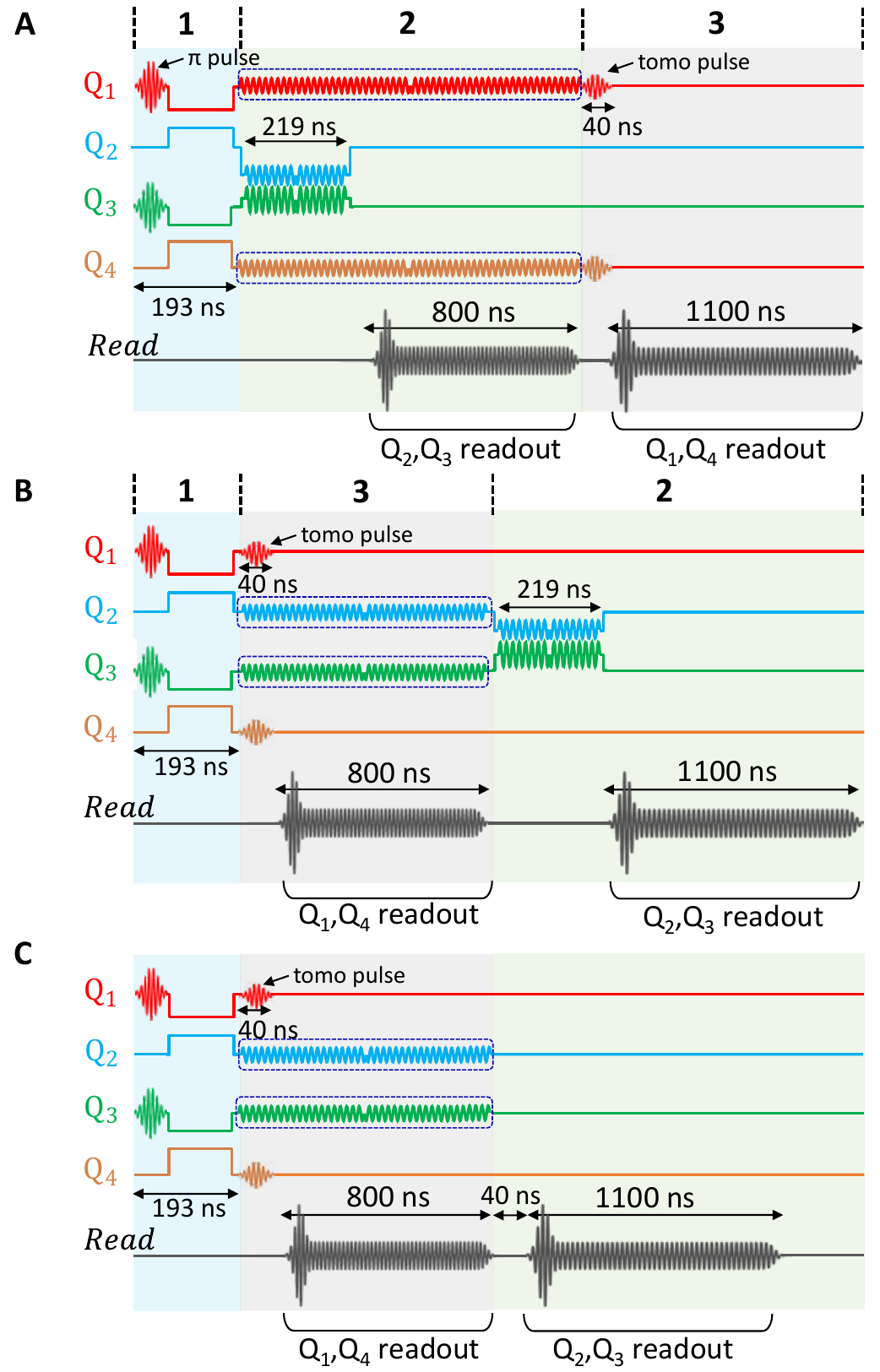}
	\caption{\label{seq12} \footnotesize{Experimental pulse sequence. (a) Pulse sequence for normal entanglement swapping. The sequence is divided into 3 successive steps: 1. entangled state preparation; 2. Bell state measurement ($Q_2$ and $Q_3$); and 3. quantum state tomography ($Q_1$ and $Q_4$). During the first step, each of the two qubit pairs $Q_{1}$-$Q_{2}$ and $Q_{3}$-$Q_{4}$ is tuned on resonance respectively for a specific time (about 153 ns for $Q_1$ and $Q_2$ and 114 ns for $Q_3$ and $Q_4$) with a rectangular pulse, realizing the $\sqrt{i\rm{SWAP}}$ gate. The $\pi$-rotation used in the first step and $\pi/2$-rotation used in the third step are realized by microwave pulses with a length of 40 ns and a full-width half maximum of 20 ns. The second step involves a dressed-state phase gate with a duration of about 219 ns and a multiplexed readout pulse of about 800 ns, during which two weak resonant drives are applied to $Q_1$ and $Q_4$ respectively to minimize the dephasing effect  (Dotted blue box). (b) Pulse sequence for delayed-choice entanglement swapping. Compared with (a), the temporal orders of the second ($Q_{2}$-$Q_{3}$ Bell state measurement) and third steps ($Q_{1}$-$Q_{4}$ quantum state tomography) are reversed. Similarly, $Q_2$ and $Q_3$ are protected from dephasing noise during the readout of $Q_1$ and $Q_4$ (Dotted blue box). (c)  Pulse sequence for delayed-choice separable-state projection. Here $Q_2$ and $Q_3$ are measured in the computational basis, which a {\it posteriori} projects $Q_1$ and $Q_4$ to a separable state.}
	}
	\end{figure}

\section{Experimental pulse sequences}
The pulse sequence for entanglement swapping in the normal temporal order is shown in Fig. \ref{seq12}(a),
where the time interval between $Q_2$-$Q_3$ readout pulse and $Q_1$-$Q_4$ readout pulse is about $40$ ns.
In our delayed-choice entanglement swapping experiment, we choose to measure
$Q_{2}$ and $Q_{3}$ in the Bell basis or in the computational basis
after joint $Q_{1}$-$Q_{4}$ measurement has been performed. The pulse sequence with the choice
of measurement in the Bell basis is shown in Fig. \ref{seq12}(b), where $Q_{2}$-$Q_{3}$ readout pulse is applied about $219$ ns after the end of $Q_{1}$-$Q_{4}$ readout pulse.
With this pulse sequence, the density matrices for $Q_{1}$ and $Q_{4}$
reconstructed from the four data subsets associated with the $Q_{2}$-$Q_{3}$
measurement outcomes $\left\vert 0_{2}\right\rangle \left\vert
0_{3}\right\rangle $, $\left\vert 0_{2}\right\rangle \left\vert
1_{3}\right\rangle $, $\left\vert 1_{2}\right\rangle \left\vert
0_{3}\right\rangle $, and $\left\vert 1_{2}\right\rangle \left\vert
1_{3}\right\rangle $ are displayed in Fig. 4(a)-(d) of the
main text, respectively. The pulse sequence with the choice of measurement
in the computational basis is shown in Fig. \ref{seq12}(c), where the dressed-state
phase gate is omitted before the detection of the states of $Q_{2}$ and $Q_{3}$,
and $Q_{2}$-$Q_{3}$ readout pulse is applied about 40 ns after the end of
$Q_{1}$-$Q_{4}$ readout pulse. In this case, the $Q_{1}$-$Q_{4}$ density matrices reconstructed
from the four subsets associated with the $Q_{2}$-$Q_{3}$ measurement
outcomes $\left\vert 0_{2}\right\rangle \left\vert 0_{3}\right\rangle $, $%
\left\vert 0_{2}\right\rangle \left\vert 1_{3}\right\rangle $, $\left\vert
1_{2}\right\rangle \left\vert 0_{3}\right\rangle $, and $\left\vert
1_{2}\right\rangle \left\vert 1_{3}\right\rangle $ are displayed in Fig.
4(e)-(h) of the main text, respectively.

As can be seen from Fig. \ref{seq12}(a), $Q_2$-$Q_3$ Bell state measurement is enabled by the combination of a dressed-state phase gate operation and multiplexed readout pulse which totally lasts about $1000$ ns. During this period of time $Q_1$ and $Q_4$ are idled and thus endure decoherence effects. To mitigate these effects, following the dephasing suppression scheme proposed in Ref. \cite{guo2018}, we apply weak continuous and resonant drives to these two qubits respectively (dotted blue box) during their idle time. As the phase is reversed in the middle of the driving pulse, the dephasing effect is reduced significantly, thus the quantum state of $Q_1$ and $Q_4$ can be well protected. The same method is employed in delayed-choice entanglement swapping experiment shown in Fig. \ref{seq12}(b), where the protection pulse is applied to $Q_2$ and $Q_3$.

In our experiment, $Q_{1}$-$Q_{4}$ joint density matrices associated with different $Q_{2}$-$Q_{3}$ measurement outcomes are reconstructed by postselection. For each of the three experiments shown in Fig. \ref{seq12}, we measure $2^4$ probabilities labelled as $P^k$=\{$P^k_{0_1 0_2 0_3 0_4}$, $P^k_{0_1 0_2 0_3 1_4}$, $P^k_{0_1 0_2 1_3 0_4}$, ..., $P^k_{1_1 1_2 1_3 1_4}$\}, where $k$ is the index of the $3^2$ tomographic operations applied to $Q_1$ and $Q_4$ before the joint readout. After readout correction, the probabilities are then sorted into four subsets, each of which is associated with one of $Q_2$-$Q_3$ measurement outcomes \{$\vert0_2\rangle \vert 0_3\rangle$, $\vert0_2\rangle \vert 1_3\rangle$, $\vert1_2 \rangle \vert0_3\rangle$, $\vert1_2\rangle \vert1_3\rangle$\}. For example, probabilities $P_{0_20_3}^k$=\{$P^k_{0_1 0_2 0_3 0_4}$, $P^k_{0_1 0_2 0_3 1_4}$, $P^k_{1_1 0_2 0_3 0_4}$, $P^k_{1_1 0_2 0_3 1_4}$\} are extracted to reconstruct the density matrix of $Q_1$ and $Q_4$ on condition that $Q_2$ and $Q_3$ are in $\vert0_2\rangle \vert0_3\rangle$ state. The same method applies to other cases.

\section{Results without readout error corrections for Bell state measurement}

\begin{figure}[!htb]
	\centering
		\includegraphics[width=0.49\textwidth,clip=True]{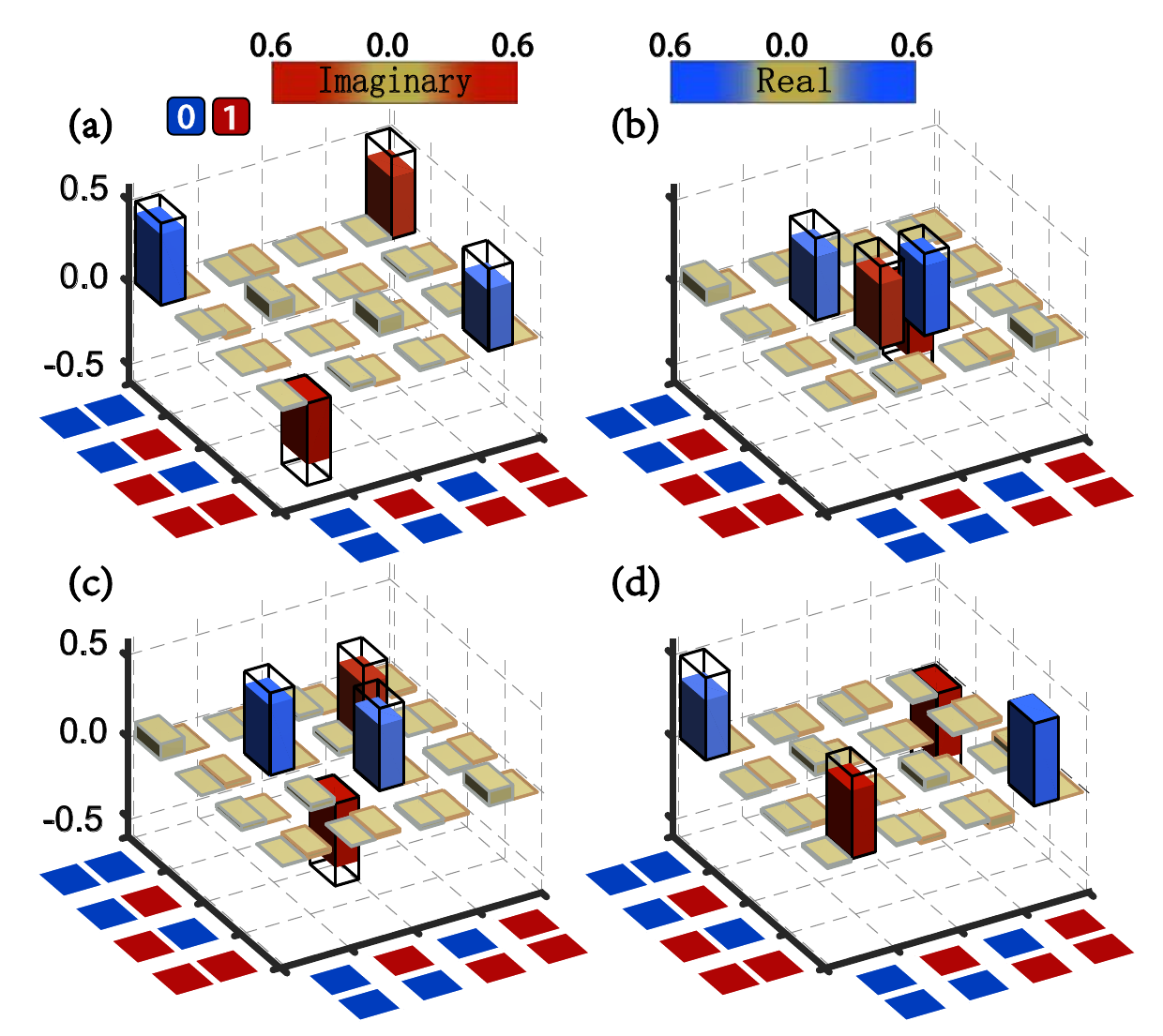}
	\caption{\label{entangleSwap_nocor_for_bell} \footnotesize{Measured $Q_{1}$-$Q_{4}$ density matrices conditional
			on the four $Q_{2}$-$Q_{3}$ measurement outcomes: (a) $\left\vert
			0_{2}\right\rangle \left\vert 0_{3}\right\rangle $;  (b) $\left\vert
			0_{2}\right\rangle \left\vert 1_{3}\right\rangle $; (c) $\left\vert
			1_{2}\right\rangle \left\vert 0_{3}\right\rangle $; (d) $\left\vert
			1_{2}\right\rangle \left\vert 1_{3}\right\rangle $. The results are obtained without performing readout error corrections for the Bell state measurement.}
	}
\end{figure}

\begin{figure*}[!htb]
	\centering
		\includegraphics[width=0.89\textwidth,clip=True]{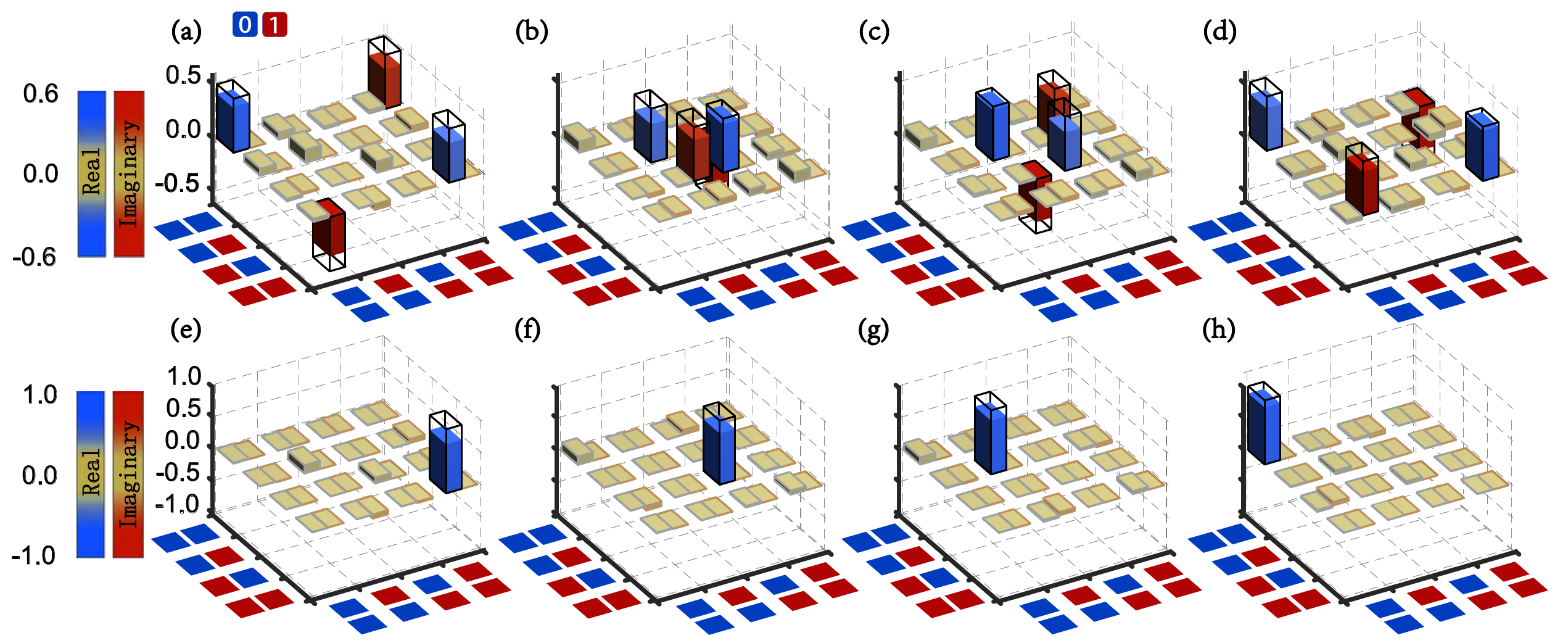}
	\caption{\label{delaychoice_nocor_for_bell} \footnotesize{Measured $Q_{1}$-$Q_{4}$ density matrices conditional
			on outcomes of delayed-choice $Q_{2}$-$Q_{3}$ measurement. (a)-(d) Results obtained from the four subsets of data correlated with the
			outcomes \{$\left\vert 0_{2}\right\rangle \left\vert 0_{3}\right\rangle $, $%
			\left\vert 0_{2}\right\rangle \left\vert 1_{3}\right\rangle $, $\left\vert
			1_{2}\right\rangle \left\vert 0_{3}\right\rangle $, $\left\vert
			1_{2}\right\rangle \left\vert 1_{3}\right\rangle $\} of $Q_{2}$-$Q_{3}$
			measurement performed after the dressed-state phase gate. (e)-(h) Results obtained from the four
			subsets of data correlated with the outcomes of the later $Q_{2}$-$Q_{3}$
			measurement without the dressed-state phase gate. All these results are obtained without performing readout error corrections for the Bell state measurement.
		}
	}
\end{figure*}

To better justify the deterministic feature of our experiment, we have also reconstructed the output density matrices without performing readout error corrections for the Bell state measurement, with the results for the normal entanglement swapping and delayed-choice entanglement swapping shown in Fig. \ref{entangleSwap_nocor_for_bell} and \ref{delaychoice_nocor_for_bell}, respectively. For the normal entanglement swapping, the fidelities of the reconstructed $Q_1$-$Q_4$ output density matrices associated with the $Q_2$-$Q_3$ Bell state measurement outcomes $%
\{\left\vert 0_{2}\right\rangle \left\vert 0_{3}\right\rangle ,\left\vert
0_{2}\right\rangle \left\vert 1_{3}\right\rangle ,\left\vert
1_{2}\right\rangle \left\vert 0_{3}\right\rangle ,\left\vert
1_{2}\right\rangle \left\vert 1_{3}\right\rangle \}$ are $0.767\pm0.008$, $0.787\pm0.009$, $0.796\pm0.010$, and $%
 0.839\pm0.010$, and the corresponding concurrences are $0.540\pm0.017$, $0.590\pm0.016$%
, $0.603\pm0.022$, and $0.691\pm0.020$, respectively. For the delayed-choice entanglement swapping, the fidelities of the four conditional $Q_1$-$Q_4$ output density matrices are $0.749\pm0.009$, $0.778\pm0.010$, $0.803\pm0.009$, and $0.828\pm0.010$, while the corresponding concurrences are $0.522\pm0.019$, $0.584\pm0.020$%
, $0.615\pm0.018$, and $0.667\pm0.018$, respectively. These results unambiguously demonstrate that $Q_1$ and $Q_4$ are deterministically projected onto an entangled state after the swapping operation.

\section{Numerical results}

The imperfections of the experimental results mainly come from the decoherence during the process. Numerical simulations with decoherence considered are performed to confirm the experimental outcomes. We use Lindblad master equation to model the state evolution under the Hamiltonian in section 1, with $T_{1,j}$ measured in experiment. In simulation, we set the pure dephasing time of $Q_j$ to be $T_{\phi,j}^{DD}$ listed in Table \ref{table1} due to the fact that a resonant continuous drive is applied to each qubit to dynamically decouple it from dephasing noises when there is a significant interval between the Bell state preparation and its state readout \cite{guo2018}. We note that although no continuous drive is applied during the initial Bell state preparation, the dephasing is also greatly depressed due to the qubit-qubit interaction, which has been discussed elsewhere \cite{mbl2018}.

Numerical simulation shows the product of $Q_1$-$Q_2$ and $Q_3$-$Q_4$ Bell states produced by swapping interactions has a fidelity of 0.979 with respect to the ideal state, which is in well agreement with the measured value of 0.971$\pm$0.009. For the dressed-state phase gate, the numerical results yield a fidelity of 0.975, slightly higher than the experimental value 0.966$\pm$0.005.

The measured probabilities of different $Q_2$-$Q_3$ basis states, and the associated $Q_{1}$-$Q_{4}$ output state fidelities and concurrences, together with the simulated results, for  the normal entanglement swapping, delayed-choice entanglement swapping, and delayed-choice separable-state projection are shown in Tables~\ref{table2}, ~\ref{table3}, and ~\ref{table4}, respectively. The experimental results are in good agreement with the numerical simulation overall. Fidelities in delayed-choice entanglement swapping are higher than those in the non-delayed case due to the fact that $Q_{1}$-$Q_{4}$ joint state is detected earlier so that the measured data is less affected by decoherence effects. The slight differences between the experimental and numerical results are partly due to the fact that $T_{\phi,j}^{DD}$, which is taken as the effective dephasing time of $Q_j$ in our simulation, does not perfectly characterize the effect of the dephasing noises under continuous driving. As demonstrated in Ref. \cite{guo2018}, the qubit's dephasing time depends not only on intensity of the applied drive but also on its effective interaction strengths with the others, which is not taken into account in the simulation. Another reason for these differences is imperfect control of the parameters of the continuous drives, whose fluctuations affect the performance of the dressed-phase gate and introduce an extra error to the protected qubits.

According to the numerical simulation, after the parallel Bell state preparations $Q_1$-$Q_4$ has a small probability of being populated in $\vert0_11_4\rangle$ when $Q_2$-$Q_3$ is projected to $\vert0_21_3\rangle$ due to the imperfect large detuning conditions, so that the elements associated with $\vert0_11_4\rangle\langle1_10_4\vert$ and $\vert1_10_4\rangle\langle0_11_4\vert$ of the corresponding projected $Q_1$-$Q_4$ density matrix have a magnitude of about $0.065$, which accounts for the calculated $Q_1$-$Q_4$ concurrence ($0.069$) corresponding to the $Q_2$-$Q_3$ output $\vert0_21_3\rangle$ under the delayed-choice separable-state projection. However, in experiment the measured magnitude of these off-diagonal elements is only about $0.02$ due to imperfect timing, to which these elements are extremely sensitive because they oscillate very fast during the parallel entangling gates. This magnitude is further reduced to about $0.007$ by the delay of projection, during which extra noises are introduced. Consequently the corresponding concurrence is too small to detect in experiment. The difference between the calculated $Q_1$-$Q_4$ concurrence and the measured result corresponding to the $Q_2$-$Q_3$ output $\vert1_20_3\rangle$, shown in Table~\ref{table4}, is due to the same reason.
\begin{table*}[!htb]
	\centering
	\begin{tabular}{c|c|c}
		\hline
		\hline
		$Q_2$-$Q_3$ probability distribution & $Q_1$-$Q_4$ Fidelity & $Q_1$-$Q_4$ Concurrence \\
		Experiment(Simulation)&Experiment (Simulation)&Experiment (Simulation)\\
		\hline
		$\vert 0_2\rangle\vert 0_3 \rangle$:~0.249$\pm$0.004 (0.250)&$\vert \Phi_{1,4}^+ \rangle$:~0.893$\pm$0.010 (0.895)&0.794$\pm$0.020 (0.794)\\
		$\vert 0_2\rangle\vert 1_3 \rangle$:~0.250$\pm$0.006 (0.257)&$\vert \Psi_{1,4}^- \rangle$:~0.879$\pm$0.010 (0.895)&0.779$\pm$0.020 (0.792)\\
		$\vert 1_2\rangle\vert 0_3 \rangle$:~0.255$\pm$0.005 (0.246)&$\vert \Psi_{1,4}^+ \rangle$:~0.872$\pm$0.011 (0.885)&0.758$\pm$0.024 (0.781)\\
		$\vert 1_2\rangle\vert 1_3 \rangle$:~0.246$\pm$0.004 (0.247)&$\vert \Phi_{1,4}^- \rangle$:~0.884$\pm$0.010 (0.893)&0.785$\pm$0.021 (0.792)\\
		\hline
		\hline
	\end{tabular}
	\caption{\label{table2} Measured and calculated probabilities of different $Q_2$-$Q_3$ basis states, and the associated $Q_1$-$Q_4$ output state fidelities and concurrences for the normal entanglement swapping.}
\end{table*}
\begin{table*}[!htb]
	\centering
	\begin{tabular}{c|c|c}
		\hline
		\hline
		$Q_2$-$Q_3$ probability distribution & $Q_1$-$Q_4$ Fidelity & $Q_1$-$Q_4$ Concurrence \\
		Experiment(Simulation)&Experiment (Simulation)&Experiment (Simulation)\\
		\hline
		$\vert 0_2\rangle\vert 0_3 \rangle$:~0.248$\pm$0.004 (0.251)&$\vert \Phi_{1,4}^+ \rangle$:~0.891$\pm$0.012 (0.908)&0.815$\pm$0.026 (0.820)\\
		$\vert 0_2\rangle\vert 1_3 \rangle$:~0.256$\pm$0.005 (0.258)&$\vert \Psi_{1,4}^- \rangle$:~0.891$\pm$0.012 (0.913)&0.816$\pm$0.024 (0.827)\\
		$\vert 1_2\rangle\vert 0_3 \rangle$:~0.248$\pm$0.006 (0.243)&$\vert \Psi_{1,4}^+ \rangle$:~0.896$\pm$0.010 (0.899)&0.806$\pm$0.022 (0.808)\\
		$\vert 1_2\rangle\vert 1_3 \rangle$:~0.248$\pm$0.006 (0.248)&$\vert \Phi_{1,4}^- \rangle$:~0.897$\pm$0.010 (0.908)&0.807$\pm$0.019 (0.821)\\
		\hline
		\hline
	\end{tabular}
	\caption{\label{table3} Measured and calculated probabilities of different $Q_2$-$Q_3$ basis states, and the associated $Q_1$-$Q_4$ output state fidelities and concurrences for the delayed-choice entanglement swapping.}
\end{table*}
\begin{table*}[!htb]
	\centering
	\begin{tabular}{c|c|c}
		\hline
		\hline
		$Q_2$-$Q_3$ probability distribution & $Q_1$-$Q_4$ Fidelity & $Q_1$-$Q_4$ Concurrence\\
		Experiment(Simulation)&Experiment (Simulation)&Experiment (Simulation)\\
		\hline
		$\vert 0_2\rangle\vert 0_3 \rangle$:~0.258$\pm$0.005 (0.258)&$\vert 1_1\rangle\vert 1_4\rangle$:~0.907$\pm$0.011 (0.932)&0.016$\pm$0.018 (0.000)\\
		$\vert 0_2\rangle\vert 1_3 \rangle$:~0.250$\pm$0.005 (0.257)&$\vert 1_1\rangle\vert 0_4\rangle$:~0.914$\pm$0.009 (0.934)&0.005$\pm$0.007 (0.069)\\
		$\vert 1_2\rangle\vert 0_3 \rangle$:~0.251$\pm$0.005 (0.245)&$\vert 0_1\rangle\vert 1_4\rangle$:~0.930$\pm$0.009 (0.946)&0.004$\pm$0.005 (0.038)\\
		$\vert 1_2\rangle\vert 1_3 \rangle$:~0.241$\pm$0.008 (0.240)&$\vert 0_1\rangle\vert 0_4\rangle$:~0.949$\pm$0.008 (0.958)&0.015$\pm$0.011 (0.000)\\
		\hline
		\hline
	\end{tabular}
	\caption{\label{table4} Measured and calculated probabilities of different $Q_2$-$Q_3$ basis states, and the associated $Q_1$-$Q_4$ output state fidelities and concurrences for the delayed-choice separable-state projection.}
\end{table*}

\section{Experimental setup}
\begin{figure*}[!htb]
	\centering
		\includegraphics[width=0.65\textwidth,clip=True]{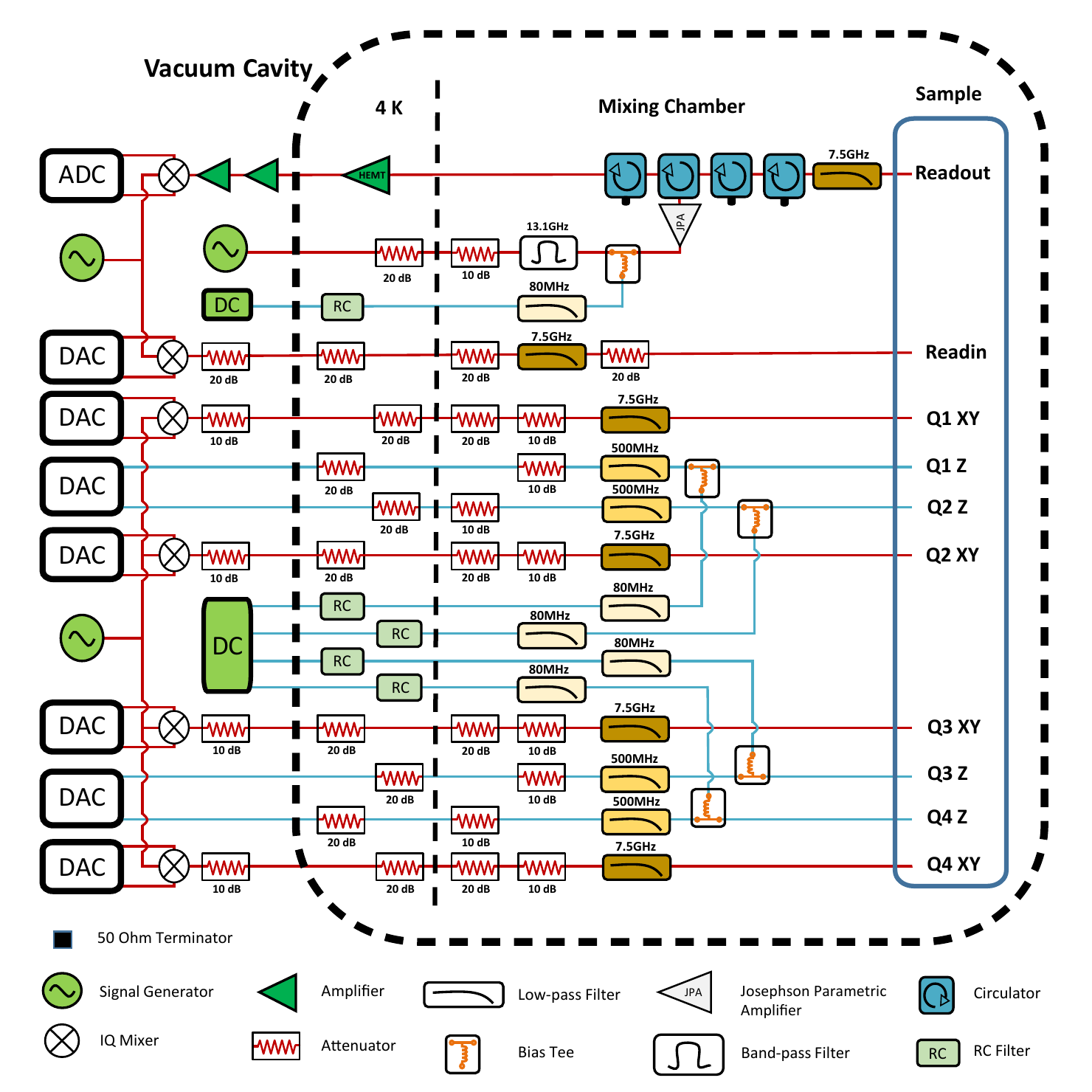}
	\caption{\label{Experimental_setup} \footnotesize{The layout of the experimental setup. Diagram showing all the control and wiring for the experimental setup. The electronics for the qubit control,
JPA control, and qubit readout are shown from the bottom to the top. The red line corresponds to the qubit XY control and measurement, while the blue line corresponds to the qubit Z control. The qubit XY control
signal is from the mixing of the signals of the two DAC channels I/Q and the microwave source, while the qubit Z control is directly produced by DAC without mixing. These two
types of qubit control realize the fast qubit flippling and frequency tuning at the nanosecond scale, respectively. Each qubit also possesses a direct-current biasing line converging to the Z line for the larger-scale frequency modulation but at a slow rate. The output signal is amplified sequentially by JPA, HEMT, room temperature amplifiers, and finally captured and demodulated by ADC. The JPA is pumped by an independent microwave signal source and its amplification band is tunable with a DC bias. Each control line goes through same attenuators and filters at each temperature stage in the dilution refrigerator to block the unwanted noises out the device.
		}
	}
\end{figure*}

Fig. \ref{Experimental_setup} shows the whole control and readout layout.
The qubit control and readout are implemented by FPGA-based
digital-to-analog converter (DAC) and analog-to-digital converter (ADC).
Each control signal output, measurement signal input and output channel is filtered
by a custom Gaussian low-pass filter with the bandwidth of 7.5 GHz
(for qubit XY control and for readin and readout control) and of 500 MHz (for qubit Z control),
respectively. The XY control signal for each qubit comes from the mixing of
the low-frequency signals sent by two independent DAC channels I/Q and
the high-frequency signal produced by the microwave source, which realizes the fast qubit
flipping at the nanosecond scale. While the qubit Z control signal is directly sent from DAC
without mixing, which enables the fast frequency tuning. The frequency can also be tuned at a slow rate through a direct-current (DC) biasing line, where the
signal generated from a low-frequency electrical source is filtered by a RC filter and a 80 MHz low-pass filter
before it is converged to qubit Z control line through a bias tee at the low
temperature stage.

The output signal from the readout feed line is amplified sequentially by the impedance-transformed
Josephson parametric amplifier (JPA), high electron mobility transistor (HEMT) and
room temperature amplifiers before it is captured and demodulated by ADC. Four cryogenic unidirectional
circulators with low insertion loss are added between JPA and the device to block the reflections
and noise emitted from the input of the HEMT at the 4K stage. The JPA is pumped by an independent
microwave signal source through the pumping line, where the signal is filtered by a 13.1 GHz bandpass filter;
the JPA's amplification band is tunable with a DC biasing line, the signal on which is filtered with
a RC filter as well as a low-pass filter before it is converged to a bias tee that connects the JPA pump line.

Every control line is balanced with some attenuators at different temperature stage in the dilution refrigerator
to prevent the unwanted noises from affecting the device.

\begin{figure*}[!htb]
	\centering
		\includegraphics[width=0.89\textwidth,clip=True]{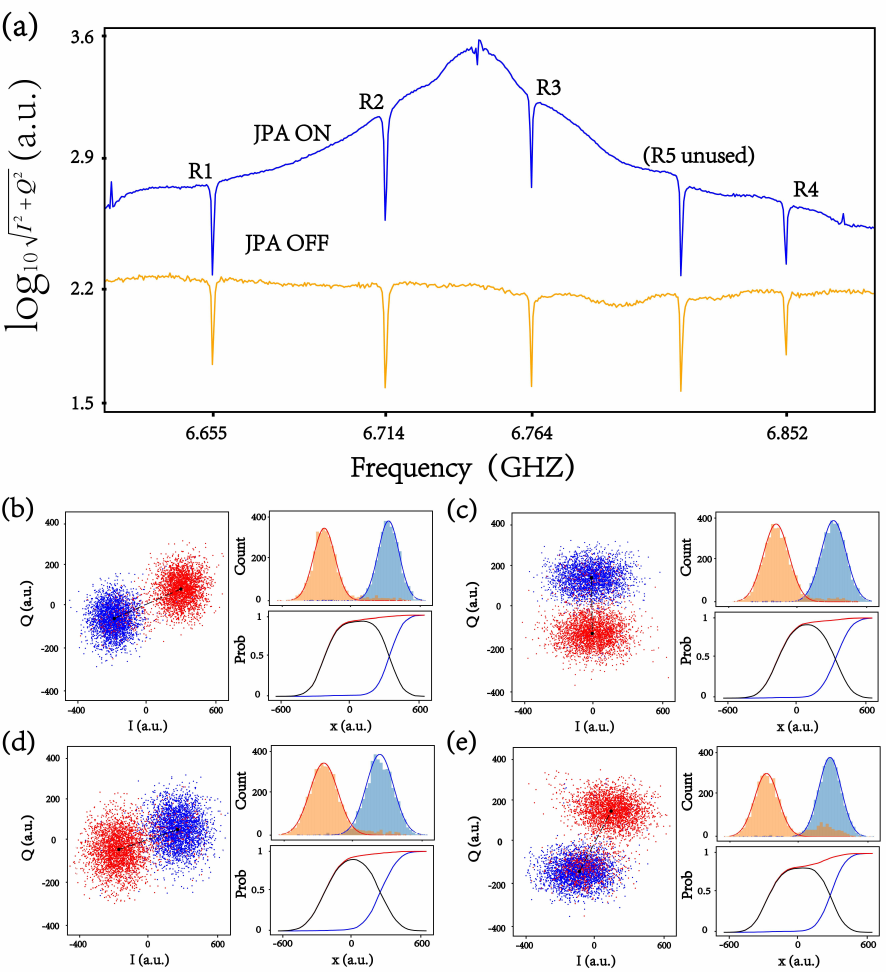}
	\caption{\label{4IQ_bandwidth} \footnotesize{(a) Signal transmission spectra through the measurement circuit (all qubits are in $|0\rangle$). The amplitudes of the demodulated signal are plotted as functions of the signal frequency, the two curves correspond to the cases when JPA is "ON" (blue) and "OFF" (orange), respectively. Five transmission dips appear at the readout resonators, which are labeled from $R_1$ to $R_5$ ($R_5$ is unused). The increased values in the longitudinal direction of the blue curve as compared to the orange one manifests the JPA's ability to amplify the signals. The measured gain ranges from 8 dBm to 22 dBm.  (b) $Q_1$, (c) $Q_2$, (d) $Q_3$, (e) $Q_4$. Left panel: Demodulated I-Q data for single-shot qubit state differentiation, where blue (red) dots are measured I-Q values when qubit is prepared in $|0\rangle$ ($|1\rangle$). The results are achieved through 3000 repetitive measurements. Top-right panel: Distribution histogram along the axis (depicted as the dotted line) in the left panel. Bottom-right panel: Measurement visibility by integrating the histogram along the dotted line axis, which has the value of about 90\%, 90\%, 88\%, and 80\% for $Q_1$, $Q_2$, $Q_3$, and $Q_4$, respectively.
		}
	}
\end{figure*}
\begin{figure}[!htb]
	\centering
	
	\includegraphics[width=0.49\textwidth,clip=True]{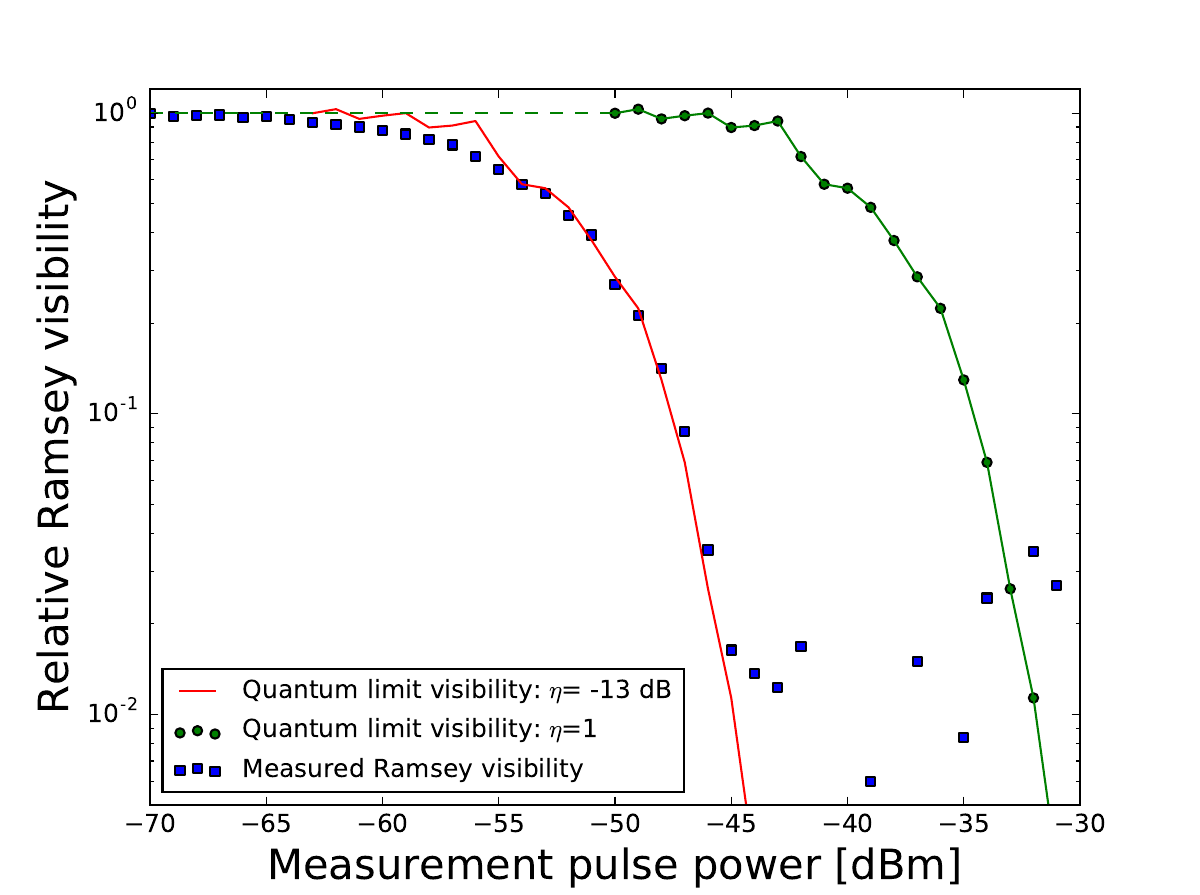}
	\caption{\label{readoutEff_q2} \footnotesize{Measurement efficiency of the system ($Q_2$ as an example). Two kinds of experiments are performed. One is the measurement of the photon induced qubit dephasing which can be charaterized by Ramsey fringe sequence, where a measurement pulse with a variable power is applied in between the two $\pi/2$-pulses, the relative fringe visibility is measured and plotted as a function of measurement pulse power (shown as blue squares). This measurement reduces the visibility of the Ramsey fringes as the applied pulse dephases the qubit. Another is the measurement of the readout distinguishability between the qubit's ground and excited state. This can be translated to a quantum limit on Ramsey fringe visibility via the relation: $\vert\rho_{10}\vert \le \frac{1}{2}exp(-\frac{SNR}{4})$ (green circles), where $\rho_{10}$ is the off-diagonal term of the density matrix of the state of the two-level system that indicates the phase coherence and SNR indicates the signal to noise ratio. The plot of this quantum limit is reshown by a left shift of it by 13 dB (red line), that passes through the points of the relative Ramsey fringe visibility.
		}
	}
\end{figure}

\section{Qubit readout}
To realize the single-shot readout demanded in this scheme, each $Q_j$ is capacitively coupled to its own readout resonator $R_j$, which can be probed by microwave pulses output from the room-temperature DAC. During the measurement, the readout pulse carrying 4 tones resonant with each $R_j$ respectively are applied through a common transmission line coupled to all $R_j$s. As the resonators are pumped with photons which dispersively interact with the qubits, the state information is encoded in the readout signal returned from the device. At the end of the microwave driving, the photons accumulated in the resonator will leak into the transmission line with a decay rate of $\kappa_j^r$ listed in Table~\ref{table1}. The returned signal is amplified by a JPA (see supplemental materials of reference \cite{10qGHZ}) to enhance the signal-to-noise ratio before it is captured and further demodulated with uniform integration weight by a room-temperature ADC. The $I$ and $Q$ values extracted from the demodulation process with each qubit in ground and excited state are illustrated in Fig. \ref{4IQ_bandwidth}(b)-(e). The results are obtained from 3000 repetitive measurements. Fig. \ref{4IQ_bandwidth}(a) shows the signal transmission spectra with JPA switched on and off. The read pulse for each qubit has the length of about 0.8 $\mu s$ or 1.1 $\mu s$, which depends on the temporal order of measurement, as can be seen from Fig. \ref{seq12}.

The measurment effeciencies are also charaterized with the method used in Ref. \cite{sank2014}. By comparing the results of two kinds of experiments: the photon-induced dephasing measured by Ramsey interference and the signal-to-noise ratio measurement, we extract the measurement efficiency, which has the values of 16 dB, 13 dB, 16 dB and 15 dB for $Q_1$, $Q_2$, $Q_3$ and $Q_4$, respectively. Fig. \ref{readoutEff_q2} shows the results of measurement efficiency by taking $Q_2$ as an example.

\newpage